\documentclass[a4paper,11pt]{article}
\usepackage{jheppub}
\usepackage{ytableau}
\usepackage{bbold}
\newcommand{\es}[2] {\begin{equation} \label{#1} \begin{split} #2 \end{split} \end{equation}}

\newcommand{\Tr}{\text{Tr}}

\newcommand{\dd}{\text{d}}

\usepackage{array}
\newcolumntype{L}{>{$}l<{$}}

\usepackage{import}
\usepackage{subfig}

\title{Holographic Equidistribution}
\author[1]{Nico Cooper}
\affiliation[1]{Department of Physics and Astronomy, University of Kentucky,\\
506 Library Drive, Lexington, KY 40506}
\emailAdd{ni.co@uky.edu}
\abstract{Hecke operators acting on modular functions arise naturally in the context of 2d conformal field theory, but in seemingly disparate areas, including permutation orbifold theories, ensembles of code CFTs, and more recently in the context of the AdS$_3$/RMT$_2$ program. We use an equidistribution theorem for Hecke operators to show that in each of these large $N$ limits, an entire heavy sector of the partition function gets integrated out, leaving only contributions from Poincar\'e series of light states. This gives an immediate holographic interpretation as a sum over semiclassical handlebody geometries. We speculate on further physical interpretations for equidistribution, including a potential ergodicity statement.}

\begin{document}
\maketitle
\flushbottom

\section{Introduction}\label{intro}
The AdS/CFT correspondence \cite{maldacena_large_1998, aharony_large_1999,witten_anti_1998} has been the most successful approach to finding a UV complete description of quantum gravity. The picture is always the same: if you want a weakly coupled theory of Einstein gravity in $d$-dimensional Anti-de Sitter, you can find a corresponding $(d-1)$-dimensional strongly coupled conformal field theory on its boundary. The original and most explicit examples have been the approaches that involve a string theory in the bulk with a supergravity limit, since these are the bulk theories where we have the most control over quantum gravitational calculations to give us a sense for how quantum gravity in any spacetime, in any dimension may work. However, there is one wrinkle: it seems that in low dimensional holography (most famously 2d JT gravity), one cannot find a \textit{single} boundary theory, but must describe the bulk in terms of an \textit{ensemble} of boundary theories \cite{saad_jt_2019}, despite the fact that quantum gravity is expected to admit individual, discrete microstates\footnote{See however \cite{barbar_holographic_2025,biggs_melonic_2026} for some low-dimensional contexts where ensemble averaging is \textit{not} strictly necessary.}. This phenomenon mirrors Wigner's calculations for decay rates of atomic nuclei needing an ensemble description, despite knowing that an atomic nucleus must have a \textit{single} Hamiltonian \cite{wigner_characteristic_1955}. 
\par
In AdS$_3$, it is still an open problem to find a \textit{single} CFT$_2$ dual to semiclassical pure Einstein gravity (with no extra matter or supersymmetry), if a single such theory even exists. On the CFT$_2$ side, this is tantamount to finding a CFT with no symmetry (global or local) but Virasoro symmetry, which has indeed never been found. Many approaches to the problem exist with weakened assumptions, like supersymmetric deformations. One such approach includes large $N$ permutation orbifolds, where $N$ copies of a boundary CFT are gauged by a rank $N$ permutation symmetry \cite{dijkgraaf_elliptic_1997,haehl_permutation_2015,belin_string_2015}. For example the D1D5 CFT \cite{callan_d-brane_1996, das_comparing_1996, dasExcitationsDstringsEntropy1996a, maldacena_black_1997} was one of the earliest contexts in which microstates of supersymmetric black holes could be computed explicitly. Permutation orbifolds have been investigated as potential holographic theories, more recently as theories admitting a supergravity point in their moduli space \cite{belin_holographic_2020,benjamin_stranger_2022,belin_mathcaln2_2020}, or as worldsheet theories of the tensionless limit of strings in AdS$_3$ \cite{eberhardt_partition_2021}. In all of these theories, one may use $SL(2,\mathbb{Z})$ \textit{Hecke operators} to express the orbifolded partition function. 
\par 
Inspired by the JT gravity picture, approaches to AdS$_3$ quantum gravity have also involved ensemble averages \cite{maloney_averaging_2020,afkhami-jeddi_free_2021,dymarsky_quantum_2021}. In fact, recent progress has been made showing that necessary points in this tentative ensemble are highly atypical \cite{belin_measure_2025,stanford_cusps_2025}, and that the averaging procedure itself is nontrivial \cite{barbar_automorphism-weighted_2025}. A well-defined moduli space where we may conduct such an average is that of Narain CFTs, classified by the group $SO(d,d;\mathbb{Z})\backslash SO(d,d;\mathbb{R})/SO(d)\times SO(d)$. This space classifies so-called \textit{Narain lattices}. In \cite{aharony_holographic_2024}, the authors found that a discrete subset of Narain theories classified by quantum error correcting codes over $
\mathbb{Z}_p\times\mathbb{Z}_p$ has likewise a discrete notion of averaging over the partition functions $Z(\tau,\bar{\tau})$ which, like permutation orbifolds, involve $SL(2,\mathbb{Z})$ Hecke operators $T_p$:
\es{avecode}{
\bar{Z}(\tau,\bar{\tau}) \sim \frac{T_p\Theta(\tau,\bar{\tau})}{p^{n-1}+1},
}
where $\Theta(\tau,\bar{\tau})$ is the lattice theta function associated to the Narain lattice of the CFT, and $n=c$ is the length of the code and the central charge of the CFT. Using a result known as \textit{equidistribution of Hecke points} \cite{goldstein_equidistribution_2003,clozel_hecke_2001}, it was conjectured that this discrete average coincides in the large $p$ limit with the full Narain average found in \cite{maloney_averaging_2020,afkhami-jeddi_free_2021}. Hecke equidistribution states that, for a square-integrable modular function $f(\tau)$,
\es{}{\lim_{N\to\infty}\left|\!\left|T_Nf(\tau,\overline{\tau})- \int_{\mathcal{F}}\frac{\dd x \dd y}{y^2}f(\tau,\overline{\tau})\right|\!\right|=0.}
This theorem states that a large $N$ Hecke operator (where $N$ is the index of the Hecke operator $T_N$) converges to a \textit{modular integral} over $\tau$. The conjectural nature of \eqref{avecode} lies in the fact that CFT partition functions are not, in general, square-integrable over the $SL(2,\mathbb{Z})$ fundamental domain of the complex upper half-plane, and square-integrability is necessary to invoke equidistribution. \par
Square-integrability happens to likewise be the obstruction to na\"ively computing the \textit{$SL(2,\mathbb{Z})$ spectral decomposition} of CFT partition functions \cite{benjamin_harmonic_2022}. There, the authors devised a way to split the partition function into square-integrable and non-square-integrable parts, while preserving modular invariance. In their notation, any CFT partition function of primaries may be written as
\es{}{
Z(\tau) = \widehat{Z}_L(\tau) + Z_{spec}(\tau),
}
where the first term is the ``modular completion of light states" which is not square-integrable, and the second term is the rest of the partition function, which \textit{is} square-integrable. Note that the modular completion in the first term may take the form of a \textit{Poincar\'e series}, denoted by  $ \,\,\widehat{} \,\,$. Using this splitting, it is clear that \textit{equidistribution of Hecke points applies to the heavy part $Z_{spec}$ of the partition function of any CFT$_2$}. So what does this have to say for any CFT where large $N$ Hecke operators appear in the partition function? 
\par
In contexts where a large $N$ Hecke operator appears in the partition function of a CFT$_2$, we show the main result: \textit{Hecke equidistribution implies that heavy terms get integrated out, leaving only terms that involve a modular completion of light states}:
\es{}{\lim_{N\to\infty}T_N Z(\tau) = T_N\widehat{Z}_L(\tau) + (\text{const.}) + \mathcal{O}(N^{-9/28}),
    }
In AdS$_3$ holography, the natural interpretation of a Poincar\'e series in a boundary CFT (or ensembles of CFTs) is that it is dual to a sum over semiclassical ``handlebody" geometries in the bulk\footnote{The Poincar\'e series is not the \textit{only} way to maintain modular invariance, but it is a \textit{minimal} choice in a way we will make precise later. See \cite{belin_universal_2026, dymarsky_tqft_2025} for a recent inclusion of non-handlebody topologies.}. Here, it arises solely as a consequence of demanding modular invariance.
\par 
The rest of the paper proceeds as follows: 
In section \ref{hecke}, we review $SL(2,\mathbb{Z})$ Hecke operators, and how they arise in the geometry of the upper-half complex plane, along with the statement of Hecke equidistribution. In section \ref{spec}, we review the setup and results of \cite{benjamin_harmonic_2022}, where the modular invariant split between light and heavy states is performed, which will allow us to implement equidistribution. Additionally, we use their results on spectral decompositions of Narain CFT partition functions to obtain explicit versions of our calculations. In section \ref{heckeapp}, we bring it all together, using Hecke equidistribution in the large $N$ limit of code CFT averges, as well as cyclic and symmetric product orbifolds, yielding new large $N$ forms of their partition functions in terms of Poincar\'e series. In each case, this gives us a way to interpret a sum over on-shell handlebody geometries. Crucially, since equidistribution has not been employed broadly in holographic contexts, these forms all at least reproduce known results/interpretations, if not presenting an \textit{entirely new large $N$ form}. Large $N$ torus partition functions of permutation orbifolds have long since been studied \cite{haehl_permutation_2015}, but here the equidistribution theorem maintains modular invariance, compared to the approximation of $\lim_{N\to\infty}T_Nf(\tau)\approx f(N\tau)$ made for such analyses. As a bonus, Hecke operators have made some other, more technical appearances in AdS$_3$ holography, which we comment on at the end of section \ref{heckeapp}, with an eye towards application of equidistribution. Finally, in section \ref{disc}, we discuss some holographic interpretations of the results of equidistribution and speculate on future directions, involving ergodic theory \cite{eskin_ergodic_2006, bost_hecke_1995} and the use of number theoretic techniques in holography \cite{perlmutter_l-function_2025}.

\section{Hecke equidistribution}\label{hecke}
% \documentclass[../main.tex]{subfiles}
% \begin{document}
In the following subsections, we will review some relevant properties of the upper-half complex plane $\mathbb{H}_2$ and its $SL(2,\mathbb{Z}$) quotient $\mathcal{F}=\mathbb{H}_2/SL(2,\mathbb{Z})$. Namely, we will review the properties of eigenfunctions of the $\mathbb{H}_2$ Laplacian, and their eigenvalues under the action of \textit{Hecke operators}. There exist many texts on modular forms in which these details may be found; we will follow the classic introduction \cite{koblitz_introduction_1993} and the text \cite{dhoker_modular_2024} for physicists.

\subsection{Some upper-half plane \texorpdfstring{$\mathbb{H}_2$}{} geometry}\label{sec:geom}
    In the fundamental domain $\mathcal{F}$ of the upper-half complex plane $\mathbb{H}_2$,
    \es{fundamentaldomain}{\mathcal{F} = \mathbb{H}_2/ SL(2,\mathbb{Z}) = \left\{ \tau = x+iy\in \mathbb{H}_2 \,\big| \,\text{Re }\tau <1/2,\, |\tau|>1
     \right\},    
    }
    we have the covariant Laplacian $\Delta_{\tau}\equiv-y^2(\partial_x^2+\partial_y^2)$, and the Weil-Petersson inner product on the space $L^2(\mathcal{F})$ of square-integrable functions:
    \es{peterssoninner}{(f,g) \equiv \int_{\mathcal{F}}\frac{\dd x\dd y}{y^2}f(\tau)\overline{g(\tau)}. 
    }
    Modular functions on $L^2(\mathcal{F})$, as they are $SL(2,\mathbb{Z})$-invariant, may be expressed in terms of a basis of modular invariant eigenfunctions of the Laplacian $\Delta_{\tau}$, just as one would decompose a spherically symmetric function into spherical harmonics. The nontrivial eigenfunctions form a discrete series ($n\in\mathbb{Z}$) and a continuous series ($s\in1/2+i\mathbb{R})$\footnote{We may analytically continue to the whole complex plane the order $s\in\mathbb{C}$ of the Eisenstein series and the more general Epstein zeta series of which they are a special case. When we consider integrating along a contour in the complex $s$ plane, we must consider the nontrivial poles, which coincide with the positive zeroes of $\zeta(2s)$, along the line $s=1/2+i\mathbb{R}$. For more on this, see chapter 3 of \cite{terras_harmonic_2013}}, the Maass cusp forms and the real analytic Eisenstein series, respectively. \par 
    The eigenvalues of $\Delta_{\tau}$ corresponding to the Maass cusp forms $\nu_{n}(\tau)$ may be parametrized as follows:
    \es{maassprop1}{\{\nu_{n\in\mathbb{Z}}\}:\Delta_{\tau} \nu_n(\tau) = \left(\frac{1}{4}+R_n^2\right)\nu_n(\tau),\quad R_n>0.}
    The Laplace eigenvalues $R_n$ of the Maass cusp forms enjoy the property of being asymptotically Poisson distributed, despite corresponding to the energy levels of a free particle propagating on a hyperbolic surface generated by an arithmetic group quotient, a system which is known to be strongly chaotic \cite{bogomolny_quantum_2003}. This emergence of Poisson statistics, usually indicative of integrability in chaotic systems on arithmetic quotient spaces, is known as \textit{arithmetic chaos} \cite{sarnak_arithmetic_1993,bogomolny_arithmetical_1997,marklof_arithmetic_2006}, and has been an important piece of evidence driving investigations into the statistical \cite{haehl_symmetries_2023, collier_wormholes_2022,ubaldo_ads_3rmt_2_2023} and number theoretic \cite{perlmutter_l-function_2025} properties of CFT partition functions.
    \par
    Next, we fix a normalization for the Weil-Petersson inner product with cusp forms by setting the value of the $n=0$ cusp form\footnote{The name ``cusp form" refers to their decay at the ``cusp" $\tau\to i\infty$.
    }:
    \es{maassprop2}{\nu_0 = \sqrt{\frac{3}{\pi}} = \text{vol}(\mathcal{F})^{-1/2}.
    }
    In addition, the $\nu_{n}(\tau)$ for $n>0$ have the property that they integrate to 0 on the real line
    \es{maassprop3}{\int_{-1/2}^{1/2}\dd x \,\nu_n(\tau) = 0,
    }
    and that they grow at most polynomially as $\tau\to i\infty$
    \es{massprop4}{\nu_n(\tau) \leq e^{-2\pi y}, \quad y\to \infty.
    }
    \par
    On the other hand, eigenfunctions corresponding to the continuous part of the spectrum of the Laplacian are the real analytic Eisenstein series $E_s(\tau)$ with $s\in1/2+i\mathbb{R}$, obeying:
    \es{eisprop1}{\{E_{s\in\frac{1}{2}+i\mathbb{R}}\}: \Delta_{\tau}E_s(\tau) = s(1-s) E_s(\tau).
    }
    At the cusp $\tau\to i\infty$, they do not decay exponentially, but rather have the following behavior:
    \es{eisprop2}{E_s(\tau) \sim y^s + \frac{\Lambda(1-s)}{\Lambda(s)}y^{1-s},}
    where $\Lambda(s) = \pi^{-s} \Gamma(s)\zeta(2s)=\Lambda(\frac{1}{2}-s)$ is known as the symmetrized Riemann zeta function.
    \par
    % Comparing the two classes of functions, note that the Eisenstein series is a class of eigenfunctions with eigenvalues directly related to its index $s$. The Maass cusp forms, on the other hand, have eigenvalues $R_n$, which may be computed in terms of Bessel functions. 
    With these functions in hand, we may compute the $SL(2,\mathbb{Z})$ spectral decomposition \cite{rankin_contributions_1939, selberg_bemerkungen_1940, zagier_rankin-selberg_1981, benjamin_harmonic_2022} of any $f(\tau)\in L^2(\mathcal{F})$:
    \es{rstrans}{\forall f(\tau)\in L^2(\mathcal{F}),\quad f(\tau)=\sum_{n=0}^{\infty}\frac{(f,\nu_n)}{(\nu_n,\nu_n)}\nu_n(\tau)+\frac{1}{4\pi i}\int_{\text{Re}s=\frac{1}{2}}\dd s (f,E_s) E_s(\tau).
    }
    In the next section, we will find that the upper-half plane Laplace eigenfunctions are simultaneously Hecke operator eigenfunctions, and their corresponding Hecke eigenvalues have fascinating statistical properties that will be of immediate use.
    
\subsection{\texorpdfstring{$SL(2,\mathbb{Z})$}{} Hecke operators}\label{sec:hecke}
    \par
    In the study of arithmetic quotients of hyperbolic spaces like $\mathbb{H}_2/SL(2,\mathbb{Z})$, there exist operators $T_n$ called \textit{Hecke operators} \cite{hecke_uber_1937} which commute mutually and  with the Laplacian. More general forms of Hecke operators and their corresponding algebras are a vast field of study on their own, and notably appear in the geometric Langlands program \cite{kapustin_electric-magnetic_2007}. In the current work, we will only need their $SL(2,\mathbb{Z})$ incarnation and their action on modular functions. \par
    There are many equivalent definitions of how Hecke operators act on modular functions (or modular forms, for that matter), but we will only need the more physically relevant action on a modular function $f(\tau)$\footnote{Our notation may be suggestive of working only over holomorphic functions of $\tau\in\mathbb{H_2}$, however Hecke operators act on holomorphic and non-holomorphic complex functions in the same way, in the sense that, for a function $f(\tau,\overline{\tau})$ Hecke operators sum over images under modular cosets acting on $\tau$ and $\overline{\tau}$ simultaneously.}. With this in mind, we'll introduce one of the more useful definitions of Hecke operators\footnote{One may also phrase the definition in terms of index $N$ lattices, as in \cite{dhoker_modular_2024}.}:
    \es{cosetheckedef}{T_Nf(\tau) = N^{k-1}\sum_{\gamma\in SL(2,\mathbb{Z})\backslash M_N}(c\tau+d)^{-k}f(\gamma \tau),
    }
    where $M_N$ is the group of $2\times2$ matrices of determinant $N$. At this point, we will now restrict to modular functions of weight 0, as this is the case for 2d CFT \textit{primary} partition functions. Finally by a parametrization of the coset $SL(2,\mathbb{Z})\backslash M_N$, we may write the sum in a much more explicit form:
    \es{heckedef}{T_N f(\tau) = \frac{1}{N}\sum_{\substack{d|N \\ d>0}}\sum_{b=0}^{d-1}f\left( \frac{N\tau + bd}{d^2} \right),
    }
    We will occasionally reduce to the case of prime $N=p$, where the action reduces to:
    \es{heckeprime}{T_pf(\tau) = \frac{1}{p}\left( f(p\tau) + \sum_{b=0}^{p-1}f\left( \frac{\tau+b}{p}\right) \right).
    }
    As a straightforward example of the composite case, $N=4$ gives:
    \es{hecke4example}{T_4f(\tau) =& \frac{1}{4}\left(f(4\tau)+f(\tau)+f\left(\frac{2\tau+1}{2}\right)+f\left(\frac{\tau}{4}\right)\right.
    \\
    &\left.+f\left(\frac{\tau+1}{4}\right)+f\left(\frac{\tau+2}{4}\right)
    +f\left(\frac{\tau+3}{4}\right) \right),
    }
    where the second and third terms are examples of the different terms that may arise from non-prime $N$.
    
    A final necessary point we emphasize is how Hecke operators act on Laplace eigenfunctions on $\mathcal{F}$, since Eisenstein series and Maass cusp forms are simultaneous eigenfunctions of the Laplacian \textit{and} Hecke operators. We may write for all $N\in\mathbb{Z}_{>0}$,
    \es{heckeeigen}{
    T_N E_s(\tau) = a_N^{(s)}E_s(\tau), \\
    T_N \nu_n (\tau) = b_N^{(n)}\nu_n(\tau),
    }
    where as before, the real analytic Eisenstein series $E_s(\tau)$ and the Maass cusp forms $\nu_n(\tau)$ are the continuous and discrete series of Laplace eigenfunctions, respectively. Namely, $a_N^{(s)}$ are defined for all $s=1/2+i\mathbb{R}$, and $b_N^{(n)}$ for all $n\in\mathbb{Z}_{>0}$. Now what are these eigenvalues? The former have an explicit form of a number theoretic flavor:\footnote{We follow the mathematical convention of a $1/N$ factor in the Hecke operator acting on a weight 0 modular function, different from the $1/\sqrt{N}$ normalization of \cite{ubaldo_ads_3rmt_2_2023,benjamin_harmonic_2022}. The main difference this makes is in the factor of $N$ in the Hecke coefficients. This means that the Hecke eigenvalues and Fourier coefficients of the Eisenstein series will technically differ by a factor of $1/\sqrt{N}$.}
    \es{eiseigenvalues}{
    a_N^{(s)} = \frac{\sigma_{2s-1}(N)}{N^{s}},
    }
    where $\sigma_s(N) \equiv \sum_{d|N}d^s$ is the analytically continued \textit{divisor sigma function}. The Maass forms, on the other had, do not have such a nice form for $b_N^{(n)}$. In fact, they have no closed form at all, and must be computed numerically. As with similar problems restricted to only numerical amenability, the reader is referred to the extensive online database \cite{the_lmfdb_collaboration_l-functions_nodate}. \par 
    What will be far more interesting to us will be what happens to $b_{N}^{(n)}$ for large $N$. This is the domain of the so-called \textit{horizontal Sato-Tate conjecture}, which (roughly) states that the $b_{N}^{(n)}$ asymptotically follow a Wigner semicircle distribution. We will later leverage these statistical-spectral properties in a similar spirit to the $AdS_3/RMT_2$ program \cite{ubaldo_ads_3rmt_2_2023,boruch_modular-invariant_2025}.
    \begin{figure}[!ht]
        \centering
        \subfloat[$Z_{\text{boson}}(\tau)$]{\includegraphics[width=0.45\linewidth]{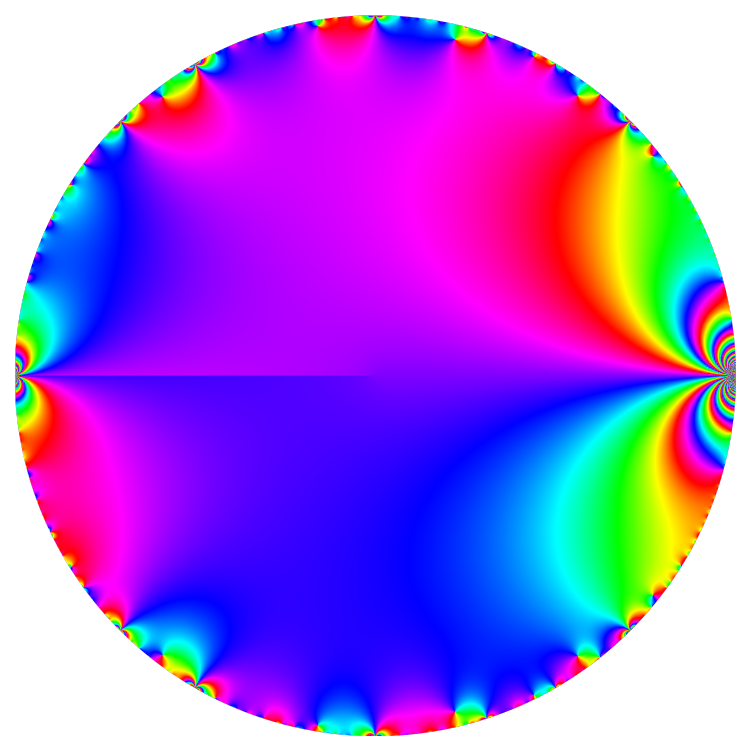}}
        \label{zboson}
        \subfloat[$T_{16}Z_{\text{boson}}(\tau)$]{\includegraphics[width=0.45\linewidth]{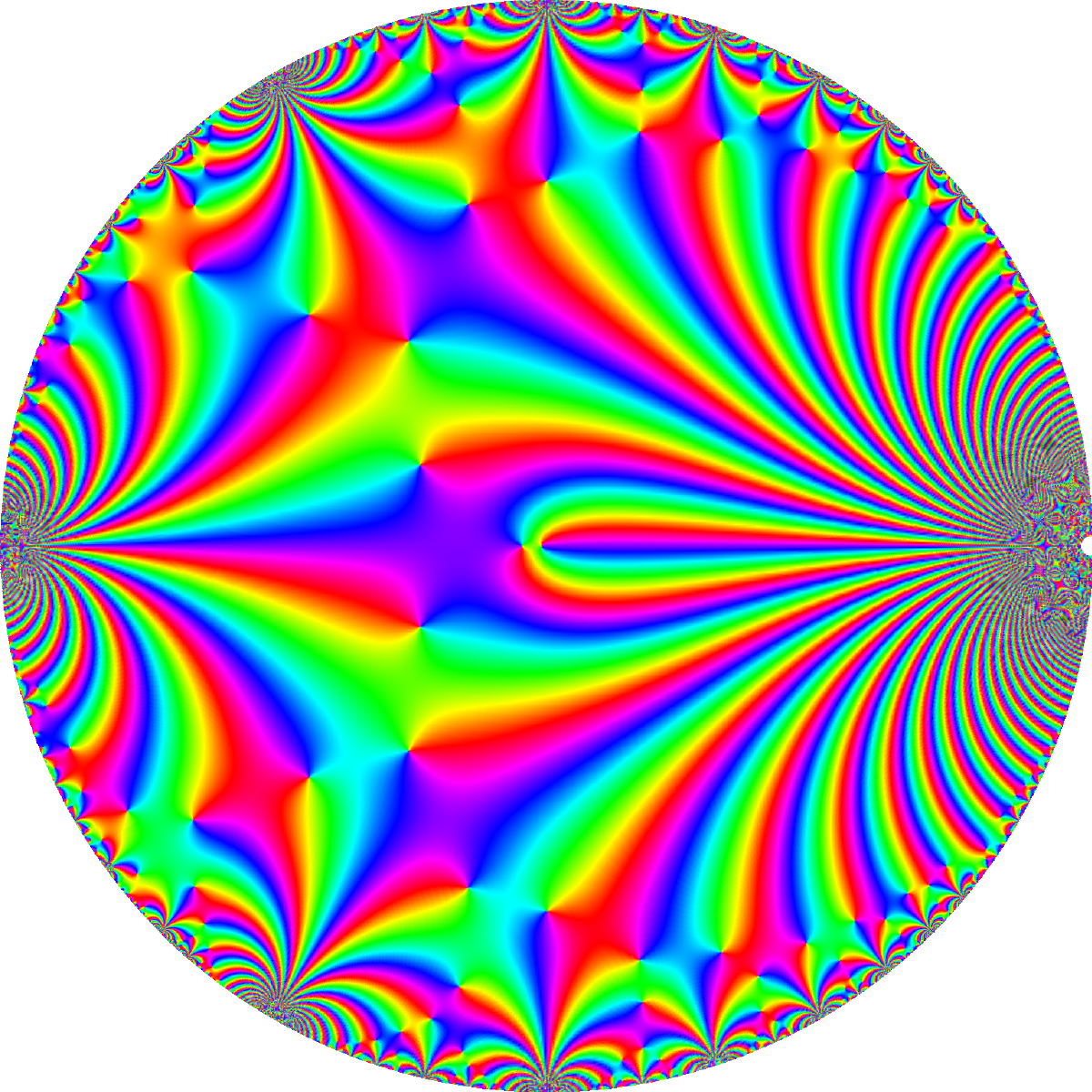}}
        \label{heckeboson16}
        \caption{Density plots in the complex $q=e^{2\pi i \tau}$ Poincar\'e disk comparing the (\textit{holomorphic}) torus partition function $Z_{\text{boson}}(\tau)$ of the \textit{chiral} $c=1$ compact boson with the Hecke operator $T_{16}$ acting on $Z_{\text{boson}}(\tau)$. Note that every pole in the plot now has a greater winding number, in particular the $T\to\infty$ divergence at $q\to1$ has its winding number multiplied by 16, since the Hecke operator multiplies the effective central charge by $N=16$. We use these plots to illustrate the effect of a Hecke operator on a modular function, but we will work exclusively with \textit{nonchiral} (nonholomorphic) partition functions otherwise.}
        \label{heckefig}
    \end{figure}
    
\subsection{Equidistribution of Hecke points}
    The current work is focused more or less on the following question: what happens when we take a large $N$ limit of a Hecke operator $T_N$ acting on a modular function $Z(\tau,\overline{\tau})$? The permutation\footnote{For a 2d CFT $\mathcal{C}$, we may tensor together $n$ copies of $\mathcal{C}$, and gauge any rank $n$ subgroup $H_n$ of the full symmetric group symmetry $S_n$. Hecke operators appear in this context for any $H_n\subseteq S_n$, we will simply use the phrase ``permutation orbifold" to refer to this broad class of theories} orbifold literature \cite{haehl_permutation_2015,belin_symmetric_2025} has investigated this question by analyzing the universal contributions to the partition functions of such theories from vacuum and low-lying states. This yields an asymptotic form of the partition function of permutation orbifold theories. However, as mentioned in the introduction, the main assumption in the literature is that $\lim_{N\to\infty}T_Nf(\tau)\approx f(N\tau)$, and the following result will give us more resolution. \par
    Here, we are interested in the more general case of \textit{whenever} Hecke operators appear for 2d CFT partition functions, and their large $N$ limit. We have the following theorem \cite{goldstein_equidistribution_2003, clozel_hecke_2001}:
    \es{equidistribution}{
    \forall\epsilon>0,\, f(\tau)\in L^2(\mathcal{F}),\quad \left|\!\left|T_Nf(\tau,\overline{\tau})- \int_{\mathcal{F}}\frac{\dd x \dd y}{y^2}f(\tau,\overline{\tau})\right|\!\right|<\left|\!\left|f(\tau,\overline{\tau})\right|\!\right|\mathcal{O}(N^{-9/28+\epsilon}),
    }
    which we may take as a result for any square-integrable modular function $f(\tau)\in L^2(\mathcal{F})$. \par
    We refer the interested reader to the original literature cited above in its native number-theoretic context. This theorem however deserves some \textit{physical} interpretation, as it tells us that an entire sector of the partition function gets integrated out as $N\to\infty$. In the rest of the paper, we provide some evidence for interpreting Hecke operators as a sort of universal modular averaging operator. Later in the discussion, we comment on some future directions for a potential ergodic interpretation of this theorem.
    \par 
    
% \end{document}

\section{Spectral decomposition}\label{spec}
% \documentclass[../main.tex]{subfiles}
% % \externaldocument{hecke}
% \begin{document}
    
\subsection{Review of \texorpdfstring{$SL(2,\mathbb{Z})$}{} spectral decomposition}
    In the previous sections, we reviewed some $\mathbb{H}_2$ geometry and the eigenfunctions of the Laplacian in the fundamental domain. What we did not emphasize is that this analysis as presented only applies to functions that are square-integrable. As noted before, CFT partition functions are not, in general, square-integrable in the fundamental domain due to contributions from sufficiently light states to the partition function. Based on work by Zagier \cite{zagier_rankin-selberg_1981}, and its subsequent application to CFT \cite{benjamin_harmonic_2022} \footnote{Earlier application of harmonic analysis to CFT partition functions was used to regulate string worldsheet one-loop integrals while maintaining modular invariance \cite{angelantonj_new_2011}.}, we may in fact apply $SL(2,\mathbb{Z})$ spectral theory to CFT partition functions in general, only assuming the requisite behavior at the cusp. Here we will review why we can do this, and how to apply spectral theory to general CFT partition functions that are in some sense ``close to square-integrable". \par
    The spectral decomposition of modular functions requires a convergent norm under the Petersson inner product defined in section \ref{sec:geom}. Merely demanding square-integrability does the job, but we can do better. A modular function is of ``slow growth" at the cusp if, for $y\to\infty$,
    \es{slowgrowth}{ Z(\tau)\sim f(y) + \text{(polynomial in $y$)},
    }
    where $f(y)$ is a sub-exponential function. For a central charge $c$ Narain CFT in particular, we may find $f(y)$ explicitly via the cuspidal behavior of the modular invariant partition function of $U(1)^c\times U(1)^c$ primaries:
    \es{naraincusp}{ Z_p(\tau) &=(y^{1/2}|\eta(\tau)|^2)^c Z(\tau)
    \\
    &\overset{y\to\infty}{\sim} y^{c/2} + \mathcal{O}(y^{c/2}e^{-2\pi \Delta_* y}),
    }
    where $\Delta_*$ is the lightest nontrivial primary dimension, and therefore $f(y)=y^{c/2}$.\par 
    In particular, the spectral decomposition of a central charge $c$ Narain CFT primary partition function takes the form \footnote{This form only applies for $c>2$, since the Rankin-Selberg transform involved in the computation is only guaranteed to converge at $c>2$ \cite{benjamin_harmonic_2022}.}:
    \es{narainspecdecomp}{ Z_{c>2}(\tau;m) =& E_{\frac{c}{2}}(\tau) + \frac{1}{4 \pi i}\int_{\text{Re } s=1/2}\dd s\, \pi^{s-\frac{c}{2}} \Gamma\left(\frac{c}{2} - s\right)\mathcal{E}_{\frac{c}{2}-s}^c (m) E_s(\tau)
    \\
    &+\sum_{n=1}^{\infty}\frac{(Z,\nu_n)}{(\nu_n,\nu_n)}\nu_n(\tau)
    +\frac{3}{\pi}\pi^{1-\frac{c}{2}}\Gamma\left(\frac{c}{2}-1\right)\mathcal{E}^c_{\frac{c}{2}-1}(m),
    }
    where $m$ are the moduli of the theory, and $\mathcal{E}_s^c(m)$ is the \textit{constrained Epstein zeta series}\footnote{We will not need these definitions here, but the constrained Epstein zeta series is defined as $\mathcal{E}_s^c(m)= \sum_{n,w\in\mathbb{Z}^c}\frac{\delta_{n\cdot w,0}}{M_{n,w}(m)}$, and $M_{n,w}(m)$ is the moduli-dependent matrix related to the $B$-field and target space metric $G$ of the Narain theory\cite{obers_eisenstein_2000,angelantonj_new_2011}.} and we have dropped the $p$ subscript on $Z$ since we will hereafter only be considering partition functions of primaries. \par
    The key features for us will be that the spectral decomposition lends itself to sharp statistical interpretations of the partition function, and more importantly that we already know how Hecke operators will act on the partition function from \eqref{heckeeigen}. 

\subsection{Splitting \texorpdfstring{$Z$}{}}\label{sec:split}
    As mentioned before, CFT partition functions are not in general $L^2(\mathcal{F})$, but we can still compute a spectral decomposition if the partition function is of slow growth at the cusp, like Narain theories. For more general classes, however, we may not even have slow growth, so is there hope at all? What comes to the rescue is the fact that partition functions are just polynomials that enumerate states, and the states that contribute divergent terms are only finitely many, since the spectrum is discretely spaced and bounded from below. \par
    In particular, the states that prevent $Z(\tau)$ from being $L^2(\mathcal{F})$ are those below the bound
    \es{btzbound}{
    \text{min}(h,\bar{h}) \leq \frac{c-c_{currents}}{24},
    }
    where $c_{currents}$ is defined as follows. For a CFT with an extended chiral algebra $\mathcal{A}$, we may inspect the high temperature behavior of the vacuum character:
    \es{}{
    \chi_{vac}^{\mathcal{A}}(y\to 0) \sim e^{\frac{i\pi}{12\tau}c_{currents}},
    }
    where we may define the effective central charge $c_{currents}$ of the chiral algebra, counting the effective current degrees of freedom. Thus, the vacuum contribution to the partition function at low temperature scales as:
    \es{}{
    Z(y\to\infty) \sim e^{\beta\left( \frac{c-c_{currents}}{12} \right)},
    }
    so the primary partition function is exponentially divergent in inverse temperature as $T\to0$. The states that contribute to this exponential divergence at low temperature are the obstruction to square-integrability that we wish to control. Above the vacuum we assume that there are only finitely many primaries in his window, so we may simply subtract them out to arrive at an $L^2(\mathcal{F})$ piece of the partition function:
    \es{zplit}{
    Z(\tau) - Z_L(\tau) \in L^2(\mathcal{F}).
    }
    Therefore the partition function may be written as a sum of the square-integrable part, and the offending light state term. However, note that neither $Z_L$ nor $Z-Z_L$ are guaranteed to be modular invariant. \par
    One of the techniques developed to maintain modular invariance in a ``minimal" way, that is, in a way that does not add new nontrivial states to the spectrum (those that break, or introduce new symmetries, for instance), is to compute a \textit{Poincar\'e series} \cite{dijkgraaf_black_2007, keller_poincare_2015, maloney_quantum_2010}, which is to simply sum over images of $SL(2,\mathbb{Z})$ which are not redundant:
    \es{zl}{
    \widehat{Z}_L(\tau) \equiv \sum_{\gamma\in \Gamma_{\infty}\backslash SL(2,\mathbb{Z})} Z_{L}(\tau_{\gamma}),
    }
    where $\gamma\in \Gamma_{\infty}\backslash SL(2,\mathbb{Z})$ is the set of  modular transformations modulo $T$-transformations. We note that this is a \textit{choice} of how to enforce modular invariance for $Z_L$. For instance, we could have summed over all modular images of $SL(2,\mathbb{Z})$. However, spin quantization already leaves the partition function invariant under $T$-transformations, hence the quotient. Other ways of constructing a modular completion exist \cite{witten_three-dimensional_2007}, but physically speaking, the Poincar\'e series is minimal in the sense that no states are added but modular images. 
    \par
    So we may define the spectrally amenable part of the partition function as:
    \es{}{
    Z(\tau) - \widehat{Z_L}(\tau) \equiv Z_{spec} \in L^2(\mathcal{F}).
    }
    In total, we may split \textit{any} CFT partition function into a piece that is square-integrable, and a piece that is a modular completion of light states. In some cases it has been shown that a boundary Poincar\'e series corresponds to a sum over bulk topological configurations with torus boundary, called \textit{handlebodies} \cite{keller_poincare_2015,maloney_averaging_2020,afkhami-jeddi_free_2021}.
\subsection{Splitting application}

    Splitting $Z(\tau)$ into square-integrable and non-square-integrable parts is a lovely tool for doing calculations with modular invariant partition functions. However, is there a physical interpretation to this split?\par 
    We have already written down the spectral decomposition for Narain theories. Which part is $Z_{spec}$, and which part is $\widehat{Z_L}$? What has been known since \cite{maloney_averaging_2020, afkhami-jeddi_free_2021} is that we may average over the Narain moduli space with Zamolodchikov measure \cite{zamolodchikov_irreversibility_1986} to arrive at a modular invariant averaged partition function of a CFT at central charge $c$ on the torus:
    \es{maloneywitten}{
    \langle Z(\tau;m)\rangle_m = \frac{E_{c/2}(\tau)}{y^{c/2} |\eta(\tau)|^{2c}}. 
    }
    The authors of \cite{di_ubaldo_ads3_2024} argue for how we may interpret the split. Note that, factoring out the $1/y^{c/2} |\eta(\tau)|^{2c}$ contribution from $U(1)^c$ descendants from the partition function\footnote{We could just factor out the Dedekind eta, but the remaining object will not be modular invariant. One may check that $1/y^{c/2} |\eta(\tau)|^{2c}$ is indeed modular invariant, so the remaining partition sum is too.  Many authors then call the remaining object a \textit{partition function of primaries}, and unless stated otherwise, \textit{we will always be working with the partition function of primaries}, as this descendant factor is the only consequence.}, this term is just the first term in the Narain CFT spectral decomposition \eqref{narainspecdecomp}. Further, $\widehat{Z}_L$ is the modular completion of states under the bound \eqref{btzbound}, for which Narain CFTs have a $U(1)^c$ chiral algebra, so $c_{currents}=c$, and this is just the modular completion of the vacuum. This is another way of defining the real analytic Eisenstein series of weight $c/2$:
    \es{vacmodcomplete}{
    \sum_{\gamma\in SL(2,\mathbb{Z})/\Gamma_{\infty}}\text{Im }(\gamma\tau)^{c/2}\sum_{\text{min}(h,\bar{h}) \leq \frac{c-c_{currents}}{24}} q_{\gamma}^{h}\overline{q_{\gamma}}^{\overline{h}} = \sum_{\gamma\in SL(2,\mathbb{Z})/\Gamma_{\infty}}\text{Im }(\gamma\tau)^{c/2} = E_{c/2}(\tau).
    }
    So already we may observe that our partition function may be written in terms of the \textit{Narain moduli} average, and $Z_{spec}$:
    \es{statsplit}{
    Z(\tau;m) = \langle Z \rangle_m + Z_{spec}(\tau;m).
    }
    and when we average over the Narain moduli space, $Z_{spec}$ averages to zero:
    \es{}{
    \langle Z_{spec}(\tau;m) \rangle_m =0,
    }
    % where we have used the fact that the modular average and he moduli space average are equivalent, due to \textit{Howe duality}, roughly the statement that the spacetime and target space Laplace eigenfunctions are related\footnote{In contexts like this, Howe duality (otherwise known as the theta correspondence) can be viewed as a generalization of the \textit{Shimura correspondence} \cite{shimura_modular_1973}, a relation between Hecke eigenspaces. For an accessible review of Howe duality and its precise application to holography from CFTs related to codes, see \cite{dymarsky_holographic_2025}, and \cite{basile_dual_2020} for a comprehensive review of Howe duality in physics.}.
    $Z_{spec}$ then captures fluctuations from the Narain moduli average, i.e. higher statistical moments, which have been studied in \cite{collier_wormholes_2022, cotler_ads_3_2020}. \par 
    Despite the fact that other CFTs will not in general have a well-defined notion of averaging over moduli, this interpretation has been extended to other CFTs \cite{collier_solving_2023,maloney_quantum_2010, chandra_semiclassical_2022}. In all, this means that the procedure of splitting the partition function as in the previous section has an interpretation as the modular \textit{average term} $\widehat{Z}_L$, and the modular \textit{fluctuation term} $Z_{spec}$. Such an interpretation has been applied to find analogues of quantum chaotic statistical methods in CFT \cite{ubaldo_ads_3rmt_2_2023}. Namely, they propose a CFT analog of the \textit{Gutzwiller trace formula} \cite{gutzwiller_periodic_1971}, which motivates the spectral interpretation of $\widehat{Z}_L + Z_{spec}$.

\section{Application of Hecke equidistribution}\label{heckeapp}
    In this section, we will apply the main result highlighted in the introduction to examples of CFT$_2$ with Hecke operators in the torus partition function. To reiterate, all we will need are the splitting procedure of \cite{benjamin_harmonic_2022} in section \ref{sec:split}, and Hecke equidistribution \cite{clozel_hecke_2001,goldstein_equidistribution_2003} in section \ref{sec:hecke}. Together, they imply that, in the large $N$ limit of the index of the Hecke operators $T_N$,
    \es{}{\lim_{N\to\infty}T_N Z(\tau) = T_N\widehat{Z}_L(\tau) + (\text{const.}) + \mathcal{O}(N^{-9/28}),
    }
    i.e. Hecke equidistribution integrates out $Z_{spec}$, the heavy part of the primary spectrum of the CFT, leaving only a modular completion of light states. Hence, the heavy sector of the seed theory with primary partition function $Z$ does not contribute in the large $N$ limit in the $T_N Z$ term.. We will explore this phenomenon in various models that exhibit near-holographic features in the large $N$ limit.

        \subsection{Code CFT averaging}\label{codeave}

        A subject of well-established study in CFT$_2$ is that of \textit{Narain} CFTs, where a theory depends on $U(1)^c\times U(1)^{\bar{c}}$ charges corresponding to momentum/winding numbers along the target space torus. These theories originally allowed for a systematic study of toroidal compactifications in string theory \cite{narain_new_1989, narain_note_1987}. Since these momenta and winding numbers are quantized, they lie on a lattice, known as a \textit{Narain lattice} $\Lambda\subset\mathbb{R}^{n,n}$, even and self-dual with respect to Lorentzian inner product. The study and classification of special lattices is a rich field on its own, yielding high-profile mathematical results using key input from string theory \cite{conway_sphere_1999}\footnote{For a lovely review of the relationships between special lattices, the Monster group, and other topics related to the numbers 8 and 24, see \cite{harvard_cmsa_quantum_matter_in_math_and_physics_john_2023}.}. Many even self-dual lattices may be related to corresponding \textit{doubly} even self-dual \textit{error correcting codes}. The simplest way to see this relationship is to take a Lorentzian lattice $\Lambda$ and reduce it modulo $p$ to obtain a code $\mathcal{C}$ over $\mathbb{Z}_p\times\mathbb{Z}_p$\footnote{This procedure for \textit{classical} codes embedded in \textit{Euclidean} lattices is known as Construction A \cite{conway_sphere_1999}. For \textit{quantum} error correcting codes embedded in \textit{Lorentzian} lattices, this is known as \textit{New} Construction A. The above paragraph and this footnote are a stand-in for the comprehensive introduction to these relationships in \cite{dymarsky_quantum_2021}. For the unfamiliar reader, we can take at face value the idea that we may label certain Narain CFTs by a \textit{code} $\mathcal{C}$ which generates the lattice $\Lambda$, as the deeper properties of error correcting codes do not enter into the physics here.}. A final piece of the correspondence that we will need is the fact that codes have a notion of a \textit{weight enumerator polynomial} $W$, which enumerates the codewords by their \textit{weight}. One might guess, and be correct, that the weight enumerator is analogous to the CFT partition function, as the partition function is a ``weight enumerator" of conformal weights. In fact, the partition function of primaries for a code CFT may indeed be written in terms of the weight enumerator $W$. 
        \par 
        Code CFTs afford us a well-defined way of performing a discrete version of the average \cite{maloney_averaging_2020, afkhami-jeddi_free_2021} over Narain CFTs. We will perform such an average at fixed $c=n$, where $n$ is the length of codewords in the code, corresponding to the central charge $c$ of the CFT. This will be a review of the average found in \cite{aharony_holographic_2024}, from which we take inspiration in using Hecke equidistribution. To begin, we review the case of $c=1$, where the ensemble is over two inequivalent codes over $\mathbb{Z}_p\times\mathbb{Z}_p$\footnote{Various generalizations abound, including code CFTs defined over nonabelian finite fields \cite{kawabata_narain_2023}, fermionic code CFTs \cite{kawabata_fermionic_2023}, and supersymmetric code CFTs \cite{kawabata_supersymmetric_2023}.}, generated by the following codewords:
        \es{c1codes}{
        (a,0)&\in\mathcal{C}_1 \rightarrow \text{compact boson with } R=R_+=\sqrt{2p}\,r
        \\
        (0,b)&\in\mathcal{C}_2 \rightarrow \text{compact boson with } R=R_-=\sqrt{2/p}\,r,}
        where $r$ is a parameter defining the embedding of the code into the Narain lattice. We have used the $\pm$ notation to denote the two distinct classes. We say ``distinct", since these CFTs obey \textit{T-duality}, the symmetry that acts as $T:R\to\frac{2}{R}$, so the $R$ moduli space at $c=1$ involves a quotient by the duality to a half-line $R\in[\sqrt{2},\infty]$, where the self-dual radius is $R=\sqrt{2}$ \footnote{Many authors alternatively choose the convention of T-duality acting as $T:R\to\frac{1}{R}$, so $R=1$ is the self-dual radius. Choosing a favorite is left as an exercise to the reader.}. When we average over these two classes CFTs, the averaged partition function is given by:
        \es{}{
        Z(R_+)+Z(R_-) = \sum_{g\in\Gamma_{0}(p)\backslash SL(2,\mathbb{Z})}\Psi_{00}(g\tau,r).
        }
        Here, $\Psi_{00}$ is the \textit{vacuum character}, defined from 
        \es{}{
        \Psi_{ab}(\tau,r) &=  \frac{1}{|\eta(\tau)|^{2n}}\sum_{n,m}e^{i\pi\tau p_L^2 - i\pi\bar{\tau}p_R^2},
        \\
        p_{L,R}=\sqrt{\frac{p}{2}}&\left( (n+a/p)r^{-1}\pm (m+b/p)r \right), \quad n,m\in\mathbb{Z},
        }
        where $a,b$ are the letters of codewords generating the code $\mathcal{C}$. This function $\Psi_{ab}$ is directly related to the weight enumerator $W$.
        
        As $p$ (roughly speaking the length of the ``alphabet" $\mathbb{Z}_p\times\mathbb{Z}_p$, and the level of the bulk Chern-Simons theory) goes to $\infty$, one recovers the full $U(1)$ symmetry of the $U(1)^c\times U(1)^c$. In the bulk path-integral representation, the average partition function at $c=1$ and when $p\to\infty$ is:
        \es{c1ave}{
        Z(R_+) + Z(R_-) = \frac{p(r+r^{-1})}{\sqrt{y}|\eta(\tau)|^2}+e^{-\mathcal{O}(p)}.
        }
        This average is divergent both as $p\to\infty$ and in the decompactification limit $r\to\infty$\footnote{This divergence is well known, see \cite{maloney_averaging_2020} for the early discussion, and \cite{barbar_global_2023,barbar_automorphism-weighted_2025} for recent interpretation involving an automorphism-weighted TQFT ensemble.}. Since this divergent average requires regularization, we will move on to the richer case of $c=2$, where the average over codes will in fact have a representation in terms of Hecke operators. 
        
        Here, Narain lattices are parametrized by the two complex moduli $t=t_1+it_2$ and $b=b_1+ib_2$. The discrete set of Narain lattices in our ensemble are classified by the following:
        \es{}{
        \{ t=\frac{k+t_0}{p} \text{  or  } t=&pt_0 , \, b=b_0\} \text{ and }
        \\
        \{t&\leftrightarrow b\},
        }
        where $k=0,\cdots ,p-1$ and $t_0,b_0$  are embedding parameters.
        \par
        At this stage we are ready to address the main issue at hand by expressing the average over $c=2$ codes with Hecke operators. In \cite{aharony_holographic_2024}, it was found that the $c=2$ average over codes may be expressed as:
        \es{c2heckeave}{
        \bar{Z}_{c=2}(\tau,t,b) = \frac{p}{2(p+1)}\left( T_p^t Z_{c=2}(\tau,t,b) + T_p^b Z_{c=2}(\tau,t,b) \right),
        }
        where $Z_{c=2}(\tau,t,b)$ is the partition function of primaries at $c=2$, proportional to the lattice theta function of the $n=2$ self-dual Narain lattice associated to a given theory:
        \es{}{
        Z_{c=2}(\tau,t,b) = y \Theta_{c=2}(\tau,t,b) = y \sum_{n,m\in\mathbb{Z}^2}e^{-\frac{\pi}{y}|\gamma(n\tau+ m)|^2-2\pi i b_1 n\wedge m},
        }
        where $\wedge$ is the Lorentzian inner product associated to the Narain lattice, and $\gamma = \mathbb{1}/\sqrt{p}$ is the metric of the toroidal target space associated to the lattice\footnote{Every lattice equivalently describes a torus by the quotient $T^{2n}=\mathbb{R}^{n,n}/\Lambda_{\mathcal{C}}$, where $\Lambda_{\mathcal{C}}$ is the Narain lattice generated by the code.}. The Hecke operators in \eqref{c2heckeave} do not act on the torus modular parameter $\tau$, but on the complex moduli $t$ and $b$, which likewise enjoy modularity. Perhaps the nicest property for our purposes is \textit{triality symmetry}---the symmetry of $Z_{c=2}$ under permuting $\tau\to t\to b\to \tau$, which allows us to write the above average as:
        \es{c2ave}{
        \bar{Z}_{c=2}(\tau,t,b) = \frac{p(T_p^{t}Z_{c=2}+T^b_{p}Z_{c=2})}{2(p+1)} = \frac{pT_p Z_{c=2}}{p+1},
        }
        where $T_p$ without the superscript denotes the ordinary action of a Hecke operator on $\tau$.
        \par
        This example is our first foray into applying Hecke equidistribution. Taking the $p\to\infty$ limit as in \eqref{c1ave}, the authors of \cite{aharony_holographic_2024} note that we may apply Hecke equidistribution to compute instead a modular integral, but this integral must be regularized as $Z_{c=2}$ is not square-integrable. However, this regularized integral was computed in \cite{dixon_moduli_1991}, from which the authors of \cite{aharony_holographic_2024} make the following conjecture:
        \es{}{
        \bar{Z}_{c=2}(\tau) = \frac{3}{\pi}\ln(p/p_0) - \frac{3}{\pi}\ln(t_2|\eta(t)|^4)-\frac{3}{\pi}\ln(b_2|\eta(b)|^4) + f(\tau) + \mathcal{O}(1/p),
        }
        where the function $f(\tau)$ is an unknown function accounting for the divergence of the modular average. This is similar to the conclusion that only the square-integrable part of the partition function of primaries modular averages away, leaving a divergent, $\tau$-dependent term. Finally, recall that $c=2$ Narain theories obey triality symmetry, so we may conclude that at this order in $1/p$, the remaining $\tau$-dependent function \textit{must} be $f(\tau) = -\frac{3}{\pi}\ln(y|\eta(\tau)|^4)$. Therefore the full averaged partition function at $c=2$ is conjectured to be
        \es{c2pave}{
        \bar{Z} = \frac{3}{\pi}\left.\frac{\ln(p/p_0) - \ln(t_2|\eta(t)|^4)-\ln(b_2|\eta(b)|^4) -\ln(y|\eta(\tau)|^4)}{y|\eta(\tau)|^4}\right|_{t=t_0,b=b_0} + \mathcal{O}(1/p).
        }
        Now, this modular integral argument is very similar to our application of Hecke equidistribution to the fluctuation term of the primary partition function. Do the approaches agree? 
        \par 
        \subsubsection*{Hecke Equidistribution for the $c=2$ Narain average}
        \par
        The spectral decompositions of Narain primary partition functions for fixed $c$ in the cases of $c=1$, $c=2$, and $c>2$ were computed by the authors of \cite{benjamin_harmonic_2022}. Here we will quote their $c=2$ result\footnote{Where $\alpha = \frac{3}{\pi}(\gamma+3\ln(4\pi)+48\zeta'(-1)-4)\approx-3.6000146$ is a numerical constant, $\Lambda(s)=\pi^{-s}\Gamma(s)\zeta(2s)$ is the symmetrized Riemann zeta function, and the $\epsilon$ notation is such that $\delta_{+}=1,$ and $\delta_{-}=-i$ distinguish between even and odd Maass cusp forms $\nu_{n}^{\epsilon}$.}:
        \es{c2narain}{
        Z_{c=2}(\tau,t,b) =& \alpha + \widehat{E}_1(\tau) + \widehat{E}_1(t) + \widehat{E}_1(b) 
        \\
        &+ \frac{1}{4\pi i }\int_{\text{Re }s=\frac{1}{2}}\dd s \,2\frac{\Lambda(s)^2}{\Lambda(1-s)}E_s(\tau)E_s(t)E_s(b)
        \\
        &+8\sum_{\epsilon=\pm}\delta_{\epsilon}\sum_{n=1}^{\infty}(\nu_n^{\epsilon},\nu_n^{\epsilon})^{-1}\nu_n^{\epsilon}(\tau)\nu_n^{\epsilon}(t)\nu_n^{\epsilon}(b),
        }
        where $\widehat{E}_1$ is the regular part of the $s=1$ Eisenstein series, defined as 
        \es{hatE}{
        \widehat{E}_1(\tau) = \lim_{s\to1}\left[ E_s(\tau) - \frac{3}{\pi(s-1)} \right].
        }
        With the spectral decomposition of the primary partition function in hand, we may reevaluate \eqref{c2ave} using only the explicit action of Hecke operators, without invoking equidistribution. The key difference between our approach and that of \cite{aharony_holographic_2024} is that we will use the splitting procedure to directly address the explicit part of the partition function that is not square-integrable, rather than introducing a regularization. This will allow us to generalize to the case of $c>2$, and give an alternative derivation without equidistribution to lend motivation to its use in general. Since \eqref{c2ave} involves only the partition function of primaries and a Hecke operator, we can plug in the spectral decomposition:
        \es{avepart}{
        \bar{Z}_{c=2}(\tau,t,b) &= \frac{p}{p+1} T_p\left(\alpha + \widehat{E}_1(\tau) + \widehat{E}_1(t) + \widehat{E}_1(b) \right)
        \\
        &+ \frac{p}{p+1}\frac{1}{4\pi i }\int_{\text{Re }s=\frac{1}{2}}\dd s \,2\frac{\Lambda(s)^2}{\Lambda(1-s)}a^{(s)}_p E_s(\tau)E_s(t)E_s(b)
        \\
        &+8\frac{p}{p+1}\sum_{\epsilon=\pm}\delta_{\epsilon}\sum_{n=1}^{\infty}(\nu_n^{\epsilon},\nu_n^{\epsilon})^{-1}b_p^{(n)}\nu_n^{\epsilon}(\tau)\nu_n^{\epsilon}(t)\nu_n^{\epsilon}(b),
        }
        where we have used the Hecke eigenvalues from \eqref{heckeeigen}, and linearity of Hecke operators to commute them with the integral over $s$. Evaluating the limit $p\to\infty$ may seem like a hefty task at first sight, but recall that we have already mentioned what to do: $Z_{spec}$ for Narain theories has a vanishing average over the Narain moduli space, which by triality symmetry is equivalent to the modular average at $c=2$, so when we take $p\to\infty$, the second and third lines ought to vanish, as these overlap terms make up $Z_{spec}$. \par 
        For the Eisenstein overlap, we may use the explicit Hecke eigenvalue, written in a particularly convenient way:
        \es{acos}{
        a_p^{(s)} = \sum_{d|p}\frac{d^{2s-1}}{p^{s}} = \sum_{d|p}\frac{1}{2\sqrt{p}}\cos\left(\omega\ln\left( d^2/p \right)\right) = \frac{1}{\sqrt{p}}\cos\left( \omega \ln p \right),
        }
        where we have juggled $s=1/2+i\omega$ with the evenness of cosine and the fact that $p$ is prime. Rewriting the integral term, one may invoke the Riemann-Lebesgue lemma to see that \textit{this integral will vanish in the large $p$ limit, as the cosine term oscillates faster as $p\to\infty$}:
        \es{eisvanish}{
        \lim_{p\to\infty}\frac{1}{4\pi \sqrt{p}}\int_{-\infty}^{\infty}\dd \omega \,2\frac{\Lambda(1/2+i\omega)^2}{\Lambda(1/2-i\omega)}\cos\left( \omega \ln p \right) E_{1/2+i\omega}(\tau)E_{1/2+i\omega}(t)E_{1/2+i\omega}(b) = 0.
        }
        Crucially, this vanishing only occurs since we are able to write the Eisenstein Fourier coefficient/Hecke eigenvalue as a highly oscillatory function in $p$, which would not be the case if $p$ were not prime. In particular, divisors of composite $N$ of order $\mathcal{O}(\sqrt{N})$ contribute an $\mathcal{O}(1)$ term in the Fourier coefficient \eqref{acos}, which would obstruct this integral from vanishing in the large $N$ limit. The fate of such terms is unclear when Hecke equidistribution should still apply for \textit{any} composite $N$. 
        \par 
        Next, consider the Maass cusp form overlap:
        \es{maassc2}{
        8\sum_{\epsilon=\pm}\delta_{\epsilon}\sum_{n=1}^{\infty}(\nu_n^{\epsilon},\nu_n^{\epsilon})^{-1}b_p^{(n)}\nu_n^{\epsilon}(\tau)\nu_n^{\epsilon}(t)\nu_n^{\epsilon}(b).
        }
        This term will be the most technically problematic, since the Maass cusp forms are erratically distributed, and little is known about them analytically. However, there is a conjecture that will be our convenient nuclear option: \textit{the horizontal Sato-Tate conjecture} for Maass cusp forms \cite{sarnak_arithmetic_1993, tate_algebraic_1965}. The conjecture roughly states that the normalized coefficients $b_p^{(n)}/b_1^{(n)}$ for fixed $n$, ordered by increasing Laplace eigenvalue $R_n$ are distributed according to the Sato-Tate distribution \cite{sarnak_statistical_1987}:
        \es{satotatedist}{
            \dd \mu_{p}(x) = 
            \begin{cases}
                \frac{(p+1)\sqrt{4-x^2}}{2\pi\left(\left(p^{1/2}+p^{-1/2}\right)^2-x^2  \right)}\dd x, & |x|<2,
                \\
                0, & \text{otherwise}.
            \end{cases}
            }
        If we take the limit as $p\to\infty$, we see the weight of the conjecture: \textit{in the large prime $p$ limit, the Hecke eigenvalues of $SL(2,\mathbb{Z})$ Maass cusp forms are distributed with respect to Wigner's semicircle distribution}, signalling random matrix statistics. This is a massive area in and of itself, but we'll just use the fact that this implies that the \textit{Maass cusp form Hecke eigenvalues are equidistributed about zero}, asymptotically obeying Wigner's surmise. So as $p\to\infty$, \eqref{maassc2} \textit{vanishes on average}:
        \es{}{
        \text{Sato-Tate conjecture }\implies \lim_{p\to\infty}8\sum_{\epsilon=\pm}\delta_{\epsilon}\sum_{n=1}^{\infty}(\nu_n^{\epsilon},\nu_n^{\epsilon})^{-1}b_p^{(s)}\nu_n^{\epsilon}(\tau)\nu_n^{\epsilon}(t)\nu_n^{\epsilon}(b)=0.
        }
        Returning to our averaged partition function \eqref{avepart}, in the large $p$ limit, we only have the first line to deal with:
        \es{}{
        \bar{Z}_{c=2}(\tau,t,b) &= \frac{p}{p+1}T_p\left(\alpha + \widehat{E}_1(\tau) + \widehat{E}_1(t) + \widehat{E}_1(b) \right).
        }
        First, a helpful fact that descends from the definition \eqref{heckedef} is that constants have Hecke eigenvalues $\frac{\sigma_1(N)}{N}$, which for $N=p$ prime is just $\frac{p+1}{p}$ so we know how to evaluate the first term. Next, we need to figure out how to handle the regular part of the $s=1$ Eisenstein series. Since the regular part subtracts out a divergent constant $\frac{\pi}{s-1}$, the Hecke operator acts on $E_{1}$ the same way as it acts on $\widehat{E}_1$: with an eigenvalue of $\frac{p+1}{p}$. As an alternative, one may invoke the \textit{Kronecker limit formula}:
        \es{kronlim}{
        \widehat{E}_1(\tau)=\frac{3}{\pi}\left( 2-24\zeta'(-1)-2\ln(4\pi) -\ln(y|\eta(\tau)|^4)\right).
        }
        Already this is starting to resemble the form of the $c=2$ average in eq.(4.40) \cite{aharony_holographic_2024}. The only additional piece is how Hecke operators act on the Dedekind eta logarithm term, which was already covered in \cite{aygunes_hecke_2010} in a more general case\footnote{The proof follows the same way as the general formulae in \cite{aygunes_hecke_2010}}, and results in the following action for integer $N$:
        \es{heckelogeta}{
        T_N\ln(y|\eta(\tau)|^4) = \frac{\sigma_{1}(N)}{N}\ln(y|\eta(\tau)|^4) - \frac{1}{N}\sum_{d|N} d\ln (d^2/N),
        }
        which in the prime $N=p$ case reduces to a form with a familiar factor, and a $ln p$ term:        
        \es{}{T_p\ln(y|\eta(\tau)|^4) = \frac{(p+1)}{p}\ln(y|\eta(\tau)|^4) - \frac{p-1}{p}\ln p.}
        Returning now to \eqref{c2ave}, the \textit{full} averaged partition function at $p\to\infty$ is
        \es{c2ave2}{
        \bar{Z}(\tau,t,b) =\frac{3}{\pi}\frac{\frac{p-1}{p}\ln p +\delta - \ln{(y|\eta(\tau)|^4)} - \ln{(t_2|\eta(t)|^4)}- \ln{(b_2|\eta(b)|^4)}}{y|\eta(\tau)|^4},
        }
        where we have packaged all of the $\tau$- and $p$-independent constants into $\delta=\gamma-3\ln 4\pi-48\zeta'(-1)+2\approx2.924$, and rewritten in terms of the log-eta terms to be able to compare to equation \eqref{c2pave} from \cite{aharony_holographic_2024}. \par
        We have now reproduced the conjectured form of the $c=2$ averaged Narain partition function, including an explicit $\ln p$ term. This term witnesses the necessity of regularization (which was subtly involved in utilizing the Kronecker limit formula), and must be thrown away in the strict $p\to\infty$ limit.

        \subsubsection*{Hecke Equidistribution for the $c>2$ Narain average}

        The previous arguments work almost identically for an average over length $n=c>2$ code theories defined over $\mathbb{Z}_p\times\mathbb{Z}_p$ in the $p\to\infty$ limit, assuming that they still take the Hecke form emphasized before. In fact this is equivalent to the bulk path integral derivation in eq.(4.17) of \cite{aharony_holographic_2024}. We may once again carry out the derivation performed above, but for the $c>2$ Narain spectral decomposition that we've met before in \eqref{narainspecdecomp}:
        \es{cg2ave}{
        \bar{Z}_{c>2}(\tau;m) &=  \frac{p^{c/2}}{p^{c-1}+1}T_p\left(E_{\frac{c}{2}}(\tau) + \frac{1}{4 \pi i}\int_{\text{Re } s=1/2}\dd s\, \pi^{s-\frac{c}{2}} \Gamma\left(\frac{c}{2} - s\right)\mathcal{E}_{\frac{c}{2}-s}^c (m) E_s(\tau)\right.
        \\
        &\left.+\sum_{n=1}^{\infty}\frac{(Z,\nu_n)}{(\nu_n,\nu_n)}\nu_n(\tau)
        +\frac{3}{\pi}\pi^{1-\frac{c}{2}}\Gamma\left(\frac{c}{2}-1\right)\mathcal{E}^c_{\frac{c}{2}-1}(m)\right),
        }
        where the combinatorial factor in front comes from equation (4.35) of \cite{aharony_holographic_2024}, and is functionally the same as the factor of $p/(p+1)$ in the $c=2$ case.
        In the $p\to\infty$ limit, this has the same behavior as the previous argument:
        \es{}{
        \bar{Z}(\tau;m) = \frac{\lim_{p\to\infty}\bar{Z}_{c>2}(\tau;m)}{y^{c}|\eta(\tau)|^{2c}} = \frac{ E_{c/2}(\tau)}{y^{c}|\eta(\tau)|^{2c}},
        }
        where the $p$-dependent factor in \eqref{cg2ave} has canceled with the Eisenstein Hecke eigenvalue as before, and the Narain average is recovered. Note that this calculation does \textit{not} involve any divergence in $p$ like those we met for $c=1$ and $c=2$, and all we have used are Hecke eigenvalues \eqref{heckeeigen} and the same argument to cancel the last two lines of \eqref{avepart}.
        \par
        Finally, it's worth commenting on a recent development in this story of averaging over CFTs classified by codes. It has been known that 2d Narain CFTs on genus $g$ tori can be related to 4d $U(1)^g$ Maxwell theory \cite{witten_s-duality_1995, verlinde_global_1995}, which is modular with respect to the orthogonal group $O(n,n,\mathbb{Z})$ on the base spacetime, and modular with respect to the symplectic group $Sp(2g,\mathbb{Z})$ on the target space. In \cite{barbar_holographic_2025}, the authors link this relationship with the recent story of Narain CFTs classified by codes. In this setup, just as we can classify some Narain CFTs by codes over discrete groups, some 4d Maxwell theories may be classified by \textit{symplectic codes}. Here there is also a holographic duality between a 5d BF theory summed over handlebodies with an ensemble of 4d Maxwell boundary conditions. There, all of our previous derivations would have analogues: \textit{orthogonal group} Hecke operators average over 4d Maxwell theories classified by symplectic codes by acting on likewise symplectic Siegel-Narain theta functions. In other words, the previous arguments would all be functionally the same, provided access to an equidistribution theorem for orthogonal group Hecke operators and $Sp(2g,\mathbb{Z)}$ spectral decompositions of 4d Maxwell partition functions. The equidistribution theorem of \cite{clozel_hecke_2001,goldstein_equidistribution_2003} generalizes to Hecke operators related to other classical groups, but to our knowledge spectral decompositions have not been performed for 4d Maxwell theories. However, since the partition functions are still Siegel-Narain theta functions, there is hope that likewise the derivation would be similar to that of \cite{benjamin_harmonic_2022}.

        \subsection{Permutation orbifolds}
        
        \subsubsection*{Cyclic Product Orbifolds}\label{cyc}
        \par
        Another way that $SL(2,\mathbb{Z})$ Hecke operators appear in 2d CFT is in the context of \textit{cyclic product orbifolds}. These are a subset of the larger class of \textit{symmetric product orbifolds}, which have been extensively studied for decades, perhaps most famously in the context of understanding black hole microstructure via \textit{D1D5 CFT} \cite{das_comparing_1996, maldacena_black_1997, callan_d-brane_1996}. One nice fact about these theories is that quantities like the partition function an $N^{\text{th}}$ cyclic product orbifold theory are related to calculating an $N^{\text{th}}$ \textit{R\'enyi entropy} in the seed theory. This was the motivation of \cite{takayanagi_free_2022}, from which we take inspiration for our calculations. We review the relationship between R\'enyi entropy and cyclic product orbifolds in Appendix \ref{renyi}. More recently, this topic has been picked up again as a potential way of understanding AdS$_3$/CFT$_2$ explicitly, in top-down and bottom-up approaches\footnote{Which direction is which is a bit fraught, and irrelevant to the current discussion, but nonetheless we point to the difference between fundamental to effective versus effective to fundamental.}. One is the approach of the tensionless limit of strings in an AdS$_3\times$S$^3\times\mathbb{T}^4$ background \cite{eberhardt_partition_2021, eberhardt_summing_2021, dei_correlators_2020}, and the other being the symmetric orbifold approach to IR-effective AdS$_3$/CFT$_2$ \cite{belin_symmetric_2025, belin_holographic_2020, haehl_permutation_2015}. 
        \par 
        In general, permutation orbifolds are performed at the level of the Hilbert space by gauging a permutation group $G$-symmetry by projecting onto the subspace of $G$-invariant states. We may represent the projection onto $G$-invariant states the following way for a torus partition function $Z$:
        \es{gaugedZ}{
        Z_{G} = \frac{1}{|G|}\sum_{\substack{g,h\in G\\gh=hg}}
        \begin{ytableau}
        \none[g] & \\
        \none & \none[h]
        \end{ytableau},
        }
        where the box notation represents setting boundary conditions on the cycles of the torus in the timelike direction twisted by $g$, and in the spacelike direction twisted by $h$, for commuting $g,h\in G$ \cite{ginsparg_applied_1988}. One way of implementing this is that we insert operators corresponding to the group elements $h$ into the trace over states in the partition function, and twist the Hilbert space itself and sum over twisted sectors by $g$:
        \es{zg}{
        Z_{G} = \frac{1}{|G|}\sum_{\substack{g,h\in G \\ gh=hg }} \text{Tr }_{\mathcal{H}_g}\left(hq^{L_0 -\frac{c}{24}}\bar{q}^{\bar{L}_0-\frac{c}{24}}\right).
        }
        The boundary conditions then act on fields as:
        \es{}{
        h\mathcal{O}(z) = \mathcal{O}(z+1) \\
        g\mathcal{O}(z) = \mathcal{O}(z+\tau).
        }
        If we consider the $N$-fold cyclic product orbifold theory Cyc($\mathcal{C}$) = $\mathcal{C}^{\otimes N}/\mathbb{Z}_N$, corresponding to tensoring $N$ copies of a seed CFT $\mathcal{C}$ and performing the above gauging procedure of the $\mathbb{Z}_N$ symmetry that permutes the copies, we may write the torus partition function as:
        \es{}{
        Z_{N,\mathbb{Z}}(\tau,\bar{\tau}) = \frac{1}{N}\sum_{g,h\in \mathbb{Z}_N}\text{Tr }\left( hq^{\mathbf{L}_0 - \frac{\mathbf{c}}{24}}\bar{q}^{\mathbf{\bar{L}}_0 - \frac{\mathbf{c}}{24}} \right),
        }
        where $\mathbf{L}_0$ and $\mathbf{c}=cN$, are the Virasoro central operator and central charge respectively of the orbifold theory, and $c$ is the central charge of the seed theory. For $N=p$ prime, an alternative way \cite{klemm_orbifolds_1990} of expressing the partition function of the orbifold theory is in terms of $SL(2,\mathbb{Z})$ Hecke operators, as promised is\footnote{All of our orbifold calculations involving a Hecke operator $T_N$ will use the normalization involving an extra factor of $1/\sqrt{N}$, i.e. \eqref{heckedef} has $1/N$ instead of $1/\sqrt{N}$, as this is the normalization used in orbifold and math literature, and the calculations go through the same way.}
        \es{primecycZ}{
        Z_{p,\mathbb{Z}}(\tau) = \frac{1}{p}Z(\tau)^p + (p-1)T_{p}Z(\tau),
        }
        where $T_p$ acts as in \eqref{heckedef}, the action and dependence on both $\tau$ and $\bar{\tau}$ are implied, and $Z(\tau)$ is the torus partition function of the seed theory. The last term on the RHS of \eqref{primecycZ} may be thought of as being responsible for the twisted sectors in the theory.
        Using the Hecke equidistribution theorem, we aim to simplify this term in the large $p$ limit, and generalize to composite $N$. \par 
        Recall that the partition function may be split as in \ref{sec:split}. As argued at the beginning of this section, the large $N$ limit for $N=p$ is amenable to the Hecke equidistribution theorem, where we may write the partition function in terms of a Poincar\'e series:
        \es{zcycp}{
        Z_{p,\mathbb{Z}} \underset{p\to\infty}{\longrightarrow} & \,\frac{1}{p}Z(\tau)^{p} + (p-1)T_p\widehat{Z}_L(\tau) + (p-1)\left(\int_{\mathcal{F}}\frac{\dd x\dd y}{y^2} Z_{\text{spec}}(\tau)+\mathcal{O}(1/p^{9/28})\right)
        \\
        =&\, \frac{1}{p}Z(\tau)^{p} + (p-1)\widehat{T_p Z_L(\tau)} + (p-1)(\text{const.} +\mathcal{O}(1/p^{9/28})),
        }
        where we have expressed the integral in $\tau$ of $Z_{\text{spec}}$ as a constant\footnote{For the rest of the $N=p$ case, we will discard this constant term along with the higher order in $p$ terms for notational convenience. We expect these terms to be physically relevant to the appropriate order of $p$, but they are as of yet unknown.}. In the last line, we have widened the hat to include the entire term, just to express the entire nontrivial twisted sector term as a Poincar\'e series, since the sum over modular images $\Gamma_{\infty}\backslash SL(2,\mathbb{Z}) $ and index $p$ conjugacy classes $\Gamma_0(p)\backslash SL(2,\mathbb{Z})$ commute. 
        \par 
        The original orbifold partition function maintains the normalization of the vacuum state even in the large $p$ limit, so this asymptotic form will only contribute a constant term that unit normalizes the vacuum degeneracy. In Appendix \ref{phase}, we review the argument of \cite{haehl_permutation_2015} that large $N$ cyclic and symmetric product orbifolds have the same low and high temperature vacuum behavior as any large $c$ CFT, guaranteeing a unique vacuum in the large $N$ limit. A new feature is reproducing the calculation with square-free Hecke operators introduced below. This derivation is more of a sanity check than anything, since \textit{any} large $c$ CFT obeys this structure, holographic/chaotic or not. 
        \par
        We may make \eqref{zcycp} more explicit by writing out the modular sums:
        \es{}{
        Z_{p\to \infty,\mathbb{Z}} =& \frac{1}{p}Z(\tau)^{p}
        + \frac{(p-1)}{p}\sum_{\substack{\min (h,\bar{h})\leq\frac{c-c_{currents}}{24} \\ \gamma\in \Gamma_{\infty}\backslash SL(2,\mathbb{Z}) \\ \gamma'\in SL(2,\mathbb{Z})\backslash M_p}} \sqrt{\text{Im}\tau_{\gamma\gamma'}}\, q^{h}_{\gamma\gamma'}\bar{q}_{\gamma\gamma'}^{\bar{h}}.
        }
        % The following commented is not quite right
        % Now, note that above we are performing a sum over two coset related to $SL(2,\mathbb{Z})$. Though not obvious, (but may be shown as in the end of chapter III of \cite{koblitz_introduction_1993}), the cosets $SL(2,\mathbb{Z})\backslash M_p$ and $\Gamma_{0}(p)\backslash SL(2,\mathbb{Z})$, where $\Gamma_{0}(p)$ is a \textit{Hecke congruence subgroup} of $SL(2,\mathbb{Z})$, are isomorphic as group actions, which will allow us a simplification in this special case. $\Gamma_{\infty}\subset\Gamma_0(p)$, so summing over equivalence classes of $SL(2,\mathbb{Z})\backslash M_p$ will be \textit{contained} in the sum over $\Gamma_{\infty}\backslash SL(2,\mathbb{Z})$. The punchline here is that summing over $SL(2,\mathbb{Z})\backslash M_p$ after $\Gamma_{\infty}\backslash SL(2,\mathbb{Z})$ will just yield a factor of $|\Gamma_{0}(p)\backslash SL(2,\mathbb{Z})|=|SL(2,\mathbb{Z})\backslash M_p|=(p+1)$:
        % \es{}{
        % Z_{p\to \infty,\mathbb{Z}} =& \frac{1}{p}Z(\tau)^{p}
        % + \frac{(p+1)(p-1)}{p}\sum_{\substack{\min (h,\bar{h})\leq\frac{c-c_{currents}}{24} \\ \gamma\in \Gamma_{\infty}\backslash SL(2,\mathbb{Z})}} \sqrt{\text{Im}\tau_{\gamma}}\, q^{h}_{\gamma}\bar{q}_{\gamma}^{\bar{h}}.
        % }
        Putting these pieces together, we have arrived at a form of the partition function at large $N=p$ that is a Poincar\'e series, plus a disconnected $p$-fold product term.\par 
        % One can show that large $N$ permutation orbifold theories have contributions to their partition function which can be interpreted holographically in terms of disconnected boundary components, and a sum over bulk handlebody geometries connecting the $N$ boundaries \cite{haehl_permutation_2015, belin_holographic_2020}. A key point of the partition function factorization puzzle is that the contribution of wormholes to the gravitational path integral has no obvious CFT analogue without ensemble averaging. Here, however, we have a context where the Poincar\'e series that is equivalent to a sum over handlebody geometries with torus boundary arises as a consequence of having a large $p$ limit for the Hecke operator $T_p$, and demanding modular invariance on the independent terms in the partition function that correspond to the average and fluctuations. 
        
        \par
        The previous argument, however, is not much more complicated than the main result highlighted in the introduction, as the partition function takes a form where part of it is simply a Hecke operator acting on the original partition function. In general, the $\mathbb{Z}_N$ cyclic product orbifold torus partition function $Z_{N,\mathbb{Z}}$ takes a more complicated, number-theoretic form in terms of the torus partition function $Z(\tau)$ of the seed theory \cite{haehl_permutation_2015}:
        \es{}{Z_{N,\mathbb{Z}}(\tau) = \frac{1}{N}\sum_{r,s=1,\cdots,N}Z\left( \frac{(N,r)}{N}\left( \frac{(N,r)}{(N,r,s)}\tau + \kappa(r,s) \right) \right)^{(N,r,s)},
        }
        where $(,)$ is the greatest common divisor of any tuple in the parentheses, and $\kappa(r,s)$ is defined as the smallest integer less than $\frac{N}{(N,r)}$ such that $\kappa(r,s)r-\frac{(N,r)s}{(N,r,s)}\equiv0$ mod $N$. As before, this reduces to \eqref{primecycZ} for prime $N=p$. This form is computationally convenient for many purposes, but for ours, we need to be able to use the appearance of Hecke operators. For entirely different reasons, expressing the full $\mathbb{Z}_N$ orbifold partition function in terms of Hecke operators was one goal of \cite{takayanagi_free_2022}. Their key observation was that a Hecke operator acting on a partition function is not ``minimally modular invariant", in the sense that there is an already modular invariant sub-term in the coset if $N$ has a perfect square divisor: the $N/a^2$-th Hecke operator, where $a^2$ is a divisor of $N$:
        \es{heckesquare}{
        \frac{1}{a^2}T_{N/a^2}Z(\tau) = \frac{1}{a^2}\frac{1}{N/a^2}\sum_{d|\frac{N}{a^2}}\sum_{j=0}^{d-1}Z\left( \frac{N/a^2}{d^2}\tau + \frac{j}{d} \right) \subset T_N Z(\tau),
        }
        where the subset denotes that the $T_{N/a^2}Z(\tau)$ sum is included in the sum in $T_N Z(\tau)$. This allows one to define the \textit{square-free Hecke operator}, which just subtracts out the square divisor term to furnish an operator that maps a modular invariant function to a \textit{minimally} modular invariant one:
        \es{sqfheckedef}{
        T_N^{sf}(\tau) = T_N Z(\tau) - \sum_{\substack{a\in\mathbb{Z}>1 \\ a^2|N}}T^{sf}_{N/a^2}Z(\tau).}
        Note that Hecke operators satisfy properties that square-free Hecke operators \textit{may not}, but the only property we will need is that square-free Hecke operators can be expressed in terms of ordinary Hecke operators.
        As an example, consider $N=36 = 2^2 3^2$. The index 36 square-free Hecke operator reads:
        \es{hecke36ex}{
        T^{sf}_{36} &= T_{36}-\frac{1}{3^2}T^{sf}_{36/3^2}-\frac{1}{2^2}T^{sf}_{36/2^2}-\frac{1}{6^2}T^{sf}_{36/6^2}
        \\
        &= T_{36}-\frac{1}{9}T^{sf}_{4}-\frac{1}{4}T^{sf}_{9}-\frac{1}{36}T_{1}
        \\
        &=T_{36}-\frac{1}{9}\left(T_{4}-\frac{1}{4}T_1\right)-\frac{1}{4}\left(T_{9}-\frac{1}{9}T_1\right)-\frac{1}{36}T_{1}
        \\
        &=T_{36}-\frac{1}{9}T_{4}-\frac{1}{4}T_{9}+\frac{1}{36}T_{1},
        }
        so the total effect can indeed be an entire re-weighting of the terms in the partition function.
        This allows us to rewrite the $\mathbb{Z}_N$ orbifold partition function in terms of square-free Hecke operators:
        \es{sqfcyc}{
        Z_{N,\mathbb{Z}}(\tau) = \sum_{d|N} \frac{\phi(N/d)}{d}T^{sf}_{N/d}\left( Z(\tau)^d \right),
        }
        where $\phi(n)$ is the Euler totient function, defined as the number of integers coprime to $n$ which are less than $n$ (defined to be 1 for $n=1$)\footnote{The Euler totient function may be rewritten in terms of he \textit{M\"obius function} $\mu(n)$ as $\phi(n)=\sum_{a|n}\mu(a)n/a$, via Dirichlet convolution \cite{terras_harmonic_2013}. More than just a number-theoretic rewriting, the M\"obius function has interesting statistical properties in that it is a \textit{pseudorandom} measure, i.e. it is a deterministic function that is almost exactly random \cite{sarnak_three_nodate}. M\"obius averages have appeared recently \cite{godet_mobius_2025} in the context of holographic dS/CFT, as an interpretation of a de Sitter Hartle-Hawking state as a M\"obius average over CFT torus partition functions, i.e. an average weighted by the M\"obius function.}. The totient function is multiplicative, and for any prime $p$, $\phi(p)=p-1$. Taken together, these will help us write partition functions for any $N=$(product of prime powers).
        
        The insight of defining square-free Hecke operators is useful for computing Renyi entropies/cyclic orbifold partition functions explicitly, but it is even more useful here for the sake of applying Hecke equidistribution. Without them, there are no Hecke operators in sight unless $N$ is prime. 
        \par
        Now that we have a Hecke operator form of the cyclic product orbifold partition function, we will now \textit{almost} apply Hecke equidistribution, but we will hit essential snags along the way that will prevent us from giving a holographic interpretation. However, cyclic product orbifolds are already known to not have a sensible holographic large $N$ limit \cite{haehl_permutation_2015}, so the composite $N$ case will serve as an exercise to prepare to look at more complex cases like symmetric product orbifolds. 
        \par 
        The first complication that arises is the fact that we are no longer able to take a large $N$ limit in the na\"ive way: the recursive definition \eqref{sqfheckedef} does not admit a large composite $N$ limit, as this depends on the square divisors of $N$, and if \textit{those} square divisors themselves have square divisors. Since we are concerned with the terms that \textit{remain} in the large $N$ limit, the remaining terms will sensitively depend on the divisors of $N$, including the contributions from square divisors that get subtracted out. To investigate this, consider the case of $N=p^2$, where there will be a single square divisor. In this case, the cyclic orbifold partition function is 
        \es{cycp2}{
        Z_{p^2,\mathbb{Z}}(\tau) &= \sum_{d|p^2} \frac{\phi(p^2/d)}{d}T^{sf}_{p^2/d}\left( Z(\tau)^d \right) 
        \\
        &= (p-1)^2T_{p^2}\left( Z(\tau) \right) - \frac{(p-1)^2}{p^2} Z(\tau) + \frac{(p-1)}{p}T_{p}\left( Z(\tau)^p \right) + \frac{1}{p^2} Z(\tau)^{p^2},
        }
        where the term proportional to $Z(\tau)$ comes from the single nontrivial square divisor of $p^2$.
        \par 
        If we attempt to perform Hecke equidistribution in \eqref{cycp2}, this will indeed go through for the first term on the RHS:
        \es{}{
         T_{p^2}Z(\tau) \underset{p\to\infty}{\longrightarrow} \widehat{T_{p^2}Z_L}.
        }
        The second place where we could use equidistribution in this setting, however, is more problematic. The immediate thought is that now the index $p$ Hecke operator term equidistributes, however the Hecke operator acts on a \textit{power} of $Z$, rather than $Z$ itself. Since powers of $Z$ are still modular invariant, we may still perform the modular invariant splitting on any power $Z^n$ in the form $Z^n = \widehat{(Z^n)}_L+(Z^n)_{spec}$. Then, equidistribution \textit{could} look like 
        \es{}{T_p(Z(\tau)^p)\underset{p\to\infty}{\longrightarrow}\widehat{T_{p}(Z^p)}_L.}
        However, the fact that $Z^p$ is $p$-dependent stands in the way of convergence of equidistribution, since $Z^p$ alone in the large $p$ limit \textit{exponentially diverges}. This divergence signals to us that there is not a physically well-defined large $N$ limit for generic cyclic product orbifolds, as the density of states has little hope of being finite. 
        \par
        In general, cyclic product orbifolds are nowhere near to being holographic theories, since first and foremost, they do not have a sparse spectrum \cite{haehl_permutation_2015, hartman_universal_2014}. Second, it has been shown that certain families $G_N\subset S_N$ in the large $N$ limit, known as \textit{oligomorphic} groups, are subgroups which admit a large $N$ limit with finite density of states. Under mild assumptions, one can show that the sizes $|G_N|$ of an oligomorphic family must grow faster than polynomially in $N$, ruling out any cyclic groups or finite products thereof \cite{belin_string_2015}. We do not claim that cyclic product orbifold theories are holographic, but they indeed are still valid as R\'enyi entropy calculations as in \cite{takayanagi_free_2022}. Thus, prime $p^{th}$ R\'enyi entropies can be related to Poincar\'e series. Likewise, these calculations serve as a warm-up to \textit{symmetric} product orbifolds that are much closer to holographic, as we will highlight in the next section.

        \subsubsection*{Symmetric Product Orbifolds}\label{sym} 

        The next type of theory we will investigate are \textit{symmetric product orbifolds}, which generalize cyclic product orbifolds in the sense that instead of gauging a cyclic group symmetry of the product theory, we gauge the \textit{entire symmetric group} $S_N$ (or any non-cyclic subgroup). 
        \par
        These theories have a much richer history than their cyclic counterparts. First and perhaps most famously, they have been studied as a model of second quantized strings, as the partition function of certain sigma model symmetric product orbifolds on a base manifold $M$ are equivalent to a second quantized string theory on $M\times S^1$ \cite{dijkgraaf_elliptic_1997}. This yields the ``DMVV formula": the grand canonical partition function of symmetric product orbifold theories with chemical potential $\rho$ conjugate to the central charge, in terms of the density of states $d(n,\bar{n})$ of the seed theory:
        \es{dmvv}{
        \mathcal{Z}(\tau) = \prod_{m>0}\prod_{\substack{n,\bar{n} \\ n-\bar{n}\equiv_m 0}} \left( 1-p^m q^{n/m}\bar{q}^{\bar{n}/m} \right)^{-d(n,\bar{n})},
        }
        where $p=e^{2\pi i \rho}$, $n,\bar{n}$ are conformal dimensions in the seed theory, and as before $q=e^{2\pi i \tau}$. This result effectively takes the symmetric product orbifold grand canonical partition function and rephrases it as a sort of Euler product, allowing one to reinterpret this grand canonical ensemble of theories as a \textit{single} free energy of a gas of D-strings in a second quantized string theory. We will not make any direct use of the DMVV formula, but more the ``left-hand side" from which \eqref{dmvv} is derived\footnote{This form is the ``original" form of the grand canonical partition function, \cite{bantay_symmetric_2000}. This formula has also appeared in holographic analyses of symmetric product orbifold theories at large $N$ \cite{keller_phase_2011, hartman_universal_2014, haehl_permutation_2015, belin_permutation_2015}, to name a few.}:
        \es{}{
        \mathcal{Z}(\tau) = \sum_{N=0}^{\infty}p^N Z_{S_{N}}(\tau) = \exp\left( \sum_{n=0}^{\infty}p^n T_n Z(\tau) \right),
        }
        where $Z(\tau) = \sum_{n,\bar{n}}d(n,\bar{n})q^n \bar{q}^{\bar{n}}$ is the partition function of the seed theory.
        \par 
        If one then expands out the exponential in an infinite series, the term multiplying $p^N$ will be the $N^{\text{th}}$ symmetric product orbifold partition function. This shows us how Hecke operators appear in the grand canonical ensemble, so if we na\"ively apply Hecke equidistribution in a similar way to before, we arrive at the statement that \textit{the terms in the grand canonical partition function of a gas of D-strings simplify at sufficiently high order}. By simplify, we mean that at sufficiently high order in the expansion of the exponential in the grand canonical partition function, the only higher contributions are terms like $p^N\widehat{T_NZ_L}$, where as before this term is a contribution of only modular images of light states. How can we interpret this simplification in this stringy context? At this point there is little that would allow us to interpret such a result, but we leave the speculation for the discussion section, where we propose a potential ergodic interpretation of equidistribution applied to physical partition functions.
        \par 
        This is as far as we can go for the grand canonical partition function, but what about individual, fixed $N$ symmetric product orbifold theories? Indeed we may use the grand canonical partition function to compute any order of partition function, but it would be more convenient to work with a single partition function at a time. For an individual symmetric product orbifold theory, we have the following partition function \cite{bantay_symmetric_2000,haehl_permutation_2015}:
        \es{snorb}{
        Z_{N,S}(\tau) = \sum_{\{m_k\}} \prod_{k=1}^{N} \frac{1}{m_k!}\left[ T_k Z(\tau) \right]^{m_k},
        }
        where the sum over $m_k$ is a sum over partitions of $N$, which run over $(m_1,\cdots,m_N)$ such that $\sum_{k=1}^{N}k m_k=N$. Partitions of $N$ are just ways of writing an integer as a sum of other integers, and their number is denoted by $p(N)$ (the number theoretic partition function). What our notation above means is the multiplicity for partitions, for instance the partition $2+1+1$ of 4 is denoted $1^2 2$, with multiplicities $m_1=2,$ $m_2=1$ and in general,
        \es{}{
        N = 1^{m_1}2^{m_2}3^{m_3}\cdots N^{m_N} \quad \text{such that }\sum_{k=1}^{N}k m_k=N.
        }
        Let's compute an example. For $N=4$, the symmetric product orbifold partition function reads 
        \es{sym4}{
        4:&\quad T_4 Z(\tau),\quad m_{1,2,3}=0,\,m_4=1,
        \\
        1\,3:&\quad Z(\tau)T_3 Z(\tau),\quad m_3,\,m_1=1,
        \\
        2^2:&\quad \frac{1}{2}(T_2Z(\tau))^2, \quad m_2=2,
        \\
        1^2 2:&\quad \frac{1}{2}Z(\tau)^2T_2 Z(\tau),\quad m_2=1,\,m_1=2,
        \\
        1^4:&\quad\frac{1}{4!}Z(\tau)^{4},\quad m_1=4,
        \\
        \implies Z_{4,S}(\tau) = T_4 Z(\tau) + Z(\tau)T_3 Z(\tau) &+ \frac{1}{2}(T_2 Z(\tau))^2 + \frac{1}{2}Z(\tau)^2 T_2 Z(\tau) + \frac{1}{4!}Z(\tau)^{4}.
        }
        This structure makes it a bit trickier to analyze a large $N$ limit, as the partitions of $N$ are not particularly ordered in a way that makes our job easier. This is why literature like \cite{haehl_permutation_2015} make the simplification that the major contribution from the Hecke operator is the term $Z(N\tau)$. How do we proceed? We must make a choice of how to consider the large $N$ limit, and therefore where equidistribution applies. 
        
        The most conservative approach would be to ask what implementation of Hecke equidistribution would result in the smallest $\mathcal{O}(N^{-9/28})$ error term. This would just mean equidistribution applies to the largest $k$ appearing in the series, i.e. the term with $T_N$ for the $N^{th}$ symmetric product orbifold, which always appears with no multiplicity. Then, we may write a large $N$ partition function with just a mild modification of the fixed $N$ case:

        \es{}{
        Z_{N\to\infty,S}(\tau)\overset{?}{=}\sum_{\{m_k\}\neq\{m_N=1\}}\prod_{k=1}^{N-1}\frac{1}{m_k!}(T_kZ(\tau))^{m_k} + \widehat{T_N Z_L}+\mathcal{O}(N^{-9/28}).
        }
        Clearly the right hand side only applies for finite $N$, so this formula is not valid in the large $N$ limit, as for any fixed $N_0$, there will always be terms with higher index Hecke operators. With this in mind, however, we may write a general form which accommodates the higher order structure by simply choosing a cutoff $N_0$ for which the $\mathcal{O}(N_0^{-9/28})$ equidistribution error is small enough:

        \es{largeNsym}{
        \lim_{N\to\infty}Z_{N,S}(\tau)\sim &\sum_{\{ m_k|m_k=0\ \forall k> N_0 \}}\prod_{k=1}^{\infty}\frac{1}{m_k!}[T_k Z(\tau)]^{m_k}
        \\
        &+\sum_{\{ m_k|m_k\neq0 \text{ for some }k\geq N_0 \}}\left( \prod_{k=1}^{ N_0} [T_k Z(\tau)]^{m_k}\right)\left( \prod_{k=N_0+1}^{\infty}\left[ \widehat{T_kZ_L(\tau)} \right] ^{m_k}\right)
        \\
        &+\mathcal{O}(N_0^{-9/28}).
        }
        In words, the first line involves terms such that, in a given set of multiplicities $\{m_k\}$ of a partition of $N$, no $m_k$ are nonzero for $k>N_0$, so no terms involve $T_kZ(\tau)$ for any $k\geq N_0$. The second line is the rest, where at least one multiplicity $m_k$ is nonzero for \textit{some} $k\geq N_0$, so a $T_kZ(\tau)$ with $k\geq N_0$ is guaranteed. For those terms $T_kZ(\tau)$ in the second line where $k\geq N_0$, equidistribution is invoked, and they are only left with a modular completion of light states $\widehat{T_kZ_L(\tau)}$, with minimal error $\mathcal{O}(N_0^{-9/28})$ . Despite the fact that both terms involve infinite products, only the second line will have infinitely many terms, as the first line is truncated to involve at most a product of $N_0$ terms.

        Consider the $N=4$ case to illustrate, as we have written out all of the terms explicitly in \eqref{sym4}. If we choose $N_0=4$, then \eqref{largeNsym} will apply equidistribution to only the term which has $T_4Z$, that is, only the first term as we have written it. If instead we chose $N_0=2$, then all terms with $T_2$ or higher will invoke equidistribution. If we had an $N\to\infty$ orbifold with $N_0=2$, then there will be a finite amount of terms involving only Hecke operators up to $T_2$, and every higher index Hecke operator will equidistribute, averaging over the heavy sectors and leaving only modular invariant, light Hecke images. 
        
        One way of phrasing this physically is that, if we impose a cutoff $N_0$ past which equidistribution is valid, then this truncates the symmetric product orbifold partition function to involve only finitely many products of disconnected boundaries and their Hecke images, and infinitely many heavy sectors which get averaged away. Due to the divisor-dependent structure of the partition function of the \textit{cyclic} product orbifold, we were only able to proceed with particular cases of $N$ in the previous subsection \ref{cyc} with ``sparse" divisors in the sense that they have a small amount of square divisors. Here however, we can gain control over the general application of equidistribution by splitting up the particular partition numbers that arise in the partition function. Namely, \eqref{largeNsym} applies for \textit{arbitrary $N$}, regardless of any special number-theoretic properties, and only needing a condition on how large the cutoff $N_0$ should be based on the equidistribution convergence rate. In the discussion, we will mention some holographic interpretations of such formulae, in terms of the bulk Chern-Simons theory characterizing the topological sector of AdS$_3$ gravity.

        \subsection{Other holographic appearance of Hecke operators}\label{sec:zstring}
        Finally, we would like to briefly investigate a more technical place in AdS$_3$ holography in which $SL(2,\mathbb{Z})$ Hecke operators appear: \textit{restoring unitarity to the Maloney-Witten-Keller regularized average.} In \cite{maloney_quantum_2010,keller_poincare_2015}, AdS$_3$ pure gravity was approached by considering the MWK partition function summed over bulk handlebody geometries:
        \es{mwk}{
        Z_{MWK}(\tau) = \sum_{\gamma\in \Gamma_{\infty}\backslash SL(2,\mathbb{Z})}\sqrt{\text{Im}(\tau_{\gamma})}|q_{\gamma}^{-\xi}(1-q_{\gamma})|^2,
        }
        where $q_{\gamma}=e^{2\pi i \tau_{\gamma}}$, and $\xi=\frac{c-1}{24}$. There are two crucial problems with this partition function: it has a continuous spectrum, and the density of states is not positive semidefinite, i.e. there are negative degeneracies. In \cite{maloney_quantum_2010,keller_poincare_2015} the authors addressed this negativity by including nonperturbative ($\mathcal{O}(1)$ in $1/c$) contributions to the partition function \eqref{mwk} that correspond to gravitational instantons arising in the large $c$ limit:
        \es{}{
        Z'_{pure}\overset{c\to\infty}{\rightarrow} Z_{MWK} + \mathcal{O}(1),
        }
        where $Z'_{pure}$ is the tentative complete partition function of AdS$_3$ pure gravity. The authors of \cite{di_ubaldo_ads3_2024} attempt to cure the negativity by adding such a tentative explicit term to the partition function of the form:
        \es{}{Z_{\text{string}}(\tau) = \sum_{\gamma\in \Gamma_{\infty}\backslash SL(2,\mathbb{Z})}\sqrt{\text{Im}(\tau_{\gamma})}(2q_{\gamma}^{\xi/4}\bar{q}_{\gamma}^{\xi/4})+\text{c.c.},
        }
        which may be interpreted as adding a modular completion a pair of single primary states with data $(\Delta_*,j_*)=\left( 2\xi,\frac{\xi}{2} \right)$, corresponding to a state with optimal spectral gap $\Delta_{*}=2\xi$. In the large $\xi$ limit, this addition restores unitarity to the spectrum.
        \par
        What does this have to do with Hecke operators? Since $Z_{\text{string}}$ is the contribution from modular images of a pair of identical states with a definite spin $j_*=\xi/2$, we may interpret this contribution as involving a Hecke operator of index $N=\xi/2$ acting on a ``primitive" partition sum\footnote{Note than an additional stipulation from \cite{di_ubaldo_ads3_2024} is that $\xi\in2\mathbb{Z}$, so the Hecke operator is well-defined here for any valid $\xi$.}:
        \es{zstring}{
        Z_{\text{string}}(\tau) = T_{\xi/2}\mathcal{Z}_{\text{string}}.
        }
        Just like before, interpretation rears its head when we are able to perform a spectral decomposition:
        \es{zstringspec}{
        Z_{\text{string}}(\tau) =& \frac{1}{4\pi i}\int_{\text{Re} s=1/2}2a_{j_*}^{(s)}\frac{\Gamma\left(\frac{\frac{1}{2}-s}{2}\right)\Gamma\left(\frac{s-\frac{1}{2}}{2}\right)}{\Lambda(s)
        \Lambda(1-s)}E^*_s(\tau)
        \\
        +&\sum_{n=1}^{\infty}b_{j_*}^{(n)}\Gamma\left(\frac{1/2-n}{2}\right)\Gamma\left(\frac{1/2+n}{2}\right)\nu_n(\tau),
        }
        with Maass form Fourier coefficients $a_N^{(s)}$ and $b_N^{(n)}$ defined as in \eqref{heckeeigen}. $\mathcal{Z}_{\text{string}}$ is then given by \eqref{zstringspec} with Fourier coefficients equal to 1, i.e. without the action of the Hecke operator.
        \par
        Recall that the Eisenstein Fourier coefficients $a_{j^*}^{(s)}=\frac{\sigma_{2s-1}(j^*)}{(j^*)^{s}}$ may be rewritten as in \eqref{acos}:
        \es{}{
        a_{j^*}^{(1/2+i\omega)} = \sum_{d|j^{*}}\frac{1}{2\sqrt{j^*}}\cos\left(\omega\ln(d^2/j^{*})\right).
        }
        If we use this to inspect the integral in \eqref{zstringspec}, and attempt to invoke Riemann-Legesgue as before, we run into a snag: \textit{$j_*=\xi/2$ is not prime.} In particular, the issue this poses in the large $\xi$ limit is that divisors of $\mathcal{O}(\sqrt{j_*})$ will have $\mathcal{O}(1)$ frequency contributions to the integral, which will not vanish at all. On the other hand, the Maass cusp form term should vanish for the same reason we had before: the Sato-Tate conjecture.
        \par
        The authors of \cite{di_ubaldo_ads3_2024} use the Sato-Tate conjecture as we have in \ref{codeave} to conclude that the Eisenstein overlap term is more involved with restoring unitarity. We may in fact use Hecke equidistribution to corroborate: since $\mathcal{Z}_{string}$ is not necessarily square-integrable, Hecke equidistribution would only integrate out \textit{part} of this unitarity restoring term. As the Eisenstein series is most responsible for breaking square-integrability, equidistribution would leave only these Eisenstein terms that prevent square-integrability of $\mathcal{Z}_{string}$. In fact we can make this a bit sharper: \textit{$Z_{\text{string}}$ restores unitarity in the large $\xi$ limit due to $\mathcal{O}(\sqrt{\xi})$ divisors in the Eisenstein overlap.} As of yet, the reason for the holographic importance of these peculiar number-theoretic contributions is still a mystery.
    
% \end{document}

\section{Discussion}\label{disc}
In studying the significance of large $N$ Hecke operators in CFT$_2$, we have arrived invariably at the appearance of Poincar\'e series. This is suggestive of the gravitational path integral involving a sum over handlebody geometries, argued to be a necessary ingredient to compute the partition function of IR effective AdS$_3$ pure gravity \cite{keller_poincare_2015, maloney_quantum_2010, dymarsky_tqft_2025, aharony_holographic_2024}. The authors of \cite{keller_poincare_2015, maloney_quantum_2010} showed that a Poincar\'e series arises as a consequence of minimally demanding modular invariance, matching the semiclassical sum over geometries in the semiclassical AdS$_3$ bulk. Here, we showed an immediate consequence of that involving Hecke equidistribution: \textit{large $N$ Hecke operators likewise demand modular invariance, implying a Poincar\'e series over light states}. This lends more evidence to the role of the Poincar\'e series in generating a semiclassical bulk theory of gravity (or a bulk theory sharing many features of pure gravity). A UV complete picture of pure AdS$_3$ quantum gravity still has its bugs, with many on-going approaches, but each approach to toy models that we have investigated have more in common than once thought. \par
Briefly, we will comment on the story of averaging. In \ref{codeave}, we reviewed the story from \cite{dymarsky_quantum_2021} of ensembles of conformal field theories classified by error correcting codes, and their large $N=p$ limit coinciding with the average over Narain moduli space. The inspiration is taken from the case of JT gravity, where it has been fairly well understood that the boundary theory involves a random matrix \textit{ensemble} \cite{saad_jt_2019}. Since this story has gained traction and even inspired approaches to de Sitter holography \cite{narovlansky_double-scaled_2023}, one may wonder when ensemble averaging becomes relevant for holographic field theories. We remain agnostic about the role of averaging in 3d quantum gravity, as there remain challenges not addressed here, but the relationship between \textit{ensemble} averaging and \textit{modular} averaging for Narain CFTs readily links these two notions \cite{dymarsky_tqft_2025}.\par
Using the Hecke equidistribution theorem, we have put forward more evidence for the statistical interpretation of the modular invariant heavy and light sectors in the CFT partition function proposed by the authors of \cite{benjamin_harmonic_2022} and further developed in \cite{ubaldo_ads_3rmt_2_2023}, lending credence to the idea that \textit{modular} averaging plays a significant role. \par
Since each of the toy model approaches considered here involve large $N$ limits of Hecke operators, this provides more evidence that the TQFT gravity picture \cite{dymarsky_tqft_2025, barbar_automorphism-weighted_2025, angelinos_abelian_2025} of summing over all bulk handlebodies, including those with vortices\footnote{This is the interpretation of terms like $\widehat{T_NZ_L}$ put forward by \cite{kames-king_lion_2024}, where they identify these terms as the connected part of the average partition function. This interpretation is in line with our calculations showing that Hecke equidistribution applies to the terms corresponding to connected covering spaces.}, is on the right track, despite the issues that remain. This is a common thread in each IR effective approach to AdS$_3$ pure gravity that we have studied. \par 
With many of these approaches having stringy interpretations \cite{di_ubaldo_ads3_2024,belin_symmetric_2025,eberhardt_partition_2021}, one may wonder if each of these theories that exhibit a form of Hecke equidistribution in the large $N$ limit are avatars of a form of averaging in strings on AdS$_3$. Since we have an interpretation of the large $N$ partition function of symmetric product orbifolds as the free energy of a gas of D-strings \cite{dijkgraaf_elliptic_1997}, it would be interesting to view this application of equidistribution as a sort of thermalization statement about strings in AdS$_3$. Alternatively, with the more recent ``long string" interpretation of the symmetric product orbifold as the worldsheet theory \cite{eberhardt_partition_2021, eberhardt_summing_2021}, perhaps we may view equidistribution as a similar thermodynamic statement about AdS-scale strings wrapping the boundary, filling out the interior of the spacetime in the large $N$ limit.
\par 
Potential interpretations abound, but since little is known about what sort of stringy interpretation that these theories may share, we will proceed with the IR effective approach: the boundary CFT partition function is to be viewed as a wavefunction of the AdS$_3$ bulk in the Chern-Simons formalism, and any action that we take on CFT partition functions should be viewed as an action on bulk wavefunctions. The on-shell sum over topologies is interpreted in this way as a sum over handlebody geometries characterized by elements of $\Gamma_{\infty}\backslash SL(2,\mathbb{Z})$, each corresponding to a particular Chern-Simons path integral. Since this is the very same Poincar\'e series we encountered in the CFT, it would be interesting to find a bulk derivation of Hecke equidistribution\footnote{This wish is partially fulfilled in \cite{aharony_holographic_2024} for the $c=2$ Narain moduli space average, as well as \cite{ubaldo_ads_3rmt_2_2023} where the use of Hecke operators is holographically justified to match Eisenstein and Maass cusp forms between the boundaries of two-sided geometries.}. We will speculate below on a potential ergodic interpretation of equidistribution, potentially having an operator-algebraic avatar in the bulk.
\par 
We end with some potential future directions.
\subsubsection*{Number-theoretic asymptotics}
    In our investigation of CFT partition functions in the large $N$ limit combined with Hecke equidistribution, there are various asymptotic formulae for number theoretic functions like the Euler totient and number theoretic partition function. What role could they have? In particular, we left off the permutation orbifolds without having an asymptotic understanding of the large $N$ partition functions that depend on the asymptotic properties of partition numbers and the Euler totient that show up in \eqref{snorb} and \eqref{sqfcyc}, respectively. \par 
    A few facts may indeed be useful here. First, the Euler totient function is bounded \cite{tenenbaum_introduction_2024}:
    \es{totient bound}{
    \phi(n)\in\left[\frac{n}{e^{\gamma}\log \log n +\frac{3}{\log\log n}},n\right]
    }
    and has an average order
    \es{totientorder}{
    \sum_{n\leq N}\phi(n) \overset{N\to\infty}{=} \frac{3}{\pi^2}N^2 + \mathcal{O}(N\ln N).
    }  
    There are close connections between the asymptotic behavior of $\phi(n)$ and the Riemann hypothesis. This may indeed suggest that the question of the role of spectral statistics in these models is deeply intertwined with one of the most famous, difficult, and famously difficult problems in mathematics (as of this work).

\subsubsection*{BTZ universality}
One motivation for this work is a recent entry into the intrigue of symmetric product orbifolds for holography. The authors of \cite{belin_symmetric_2025} claim that large $N$ symmetric product orbifolds are near-holographic in the following sense: at large $N$, thermal two-point correlators in the untwisted sector of a symmetric product orbifold theory are indistinguishable from BTZ:
\es{}{
\langle\boldsymbol{O}_{Sym^N}(t_E,\phi),\boldsymbol{O}_{Sym^N}(0)\rangle_{\beta} \overset{N\to\infty}{=} \langle\boldsymbol{O}_{BTZ}(t_E,\phi),\boldsymbol{O}_{BTZ}(0)\rangle_{\beta},
}
in both the high and low temperature regimes. In some sense this means that symmetric product orbifolds try to mimic a BTZ black hole, despite aforementioned reasons that symmetric product orbifold theories are not holographic to a weakly coupled theory of pure AdS$_3$ gravity (but \textit{are} holographic to string theory in certain cases). It would be nice to see if Hecke equidistribution may be related more explicitly to these exotic properties of symmetric orbifold correlators.

    \begin{figure}[!ht]
    \centering
        \subfloat[$T_2Z(\tau)$]{\label{hecke2}\includegraphics[width=0.35\linewidth]{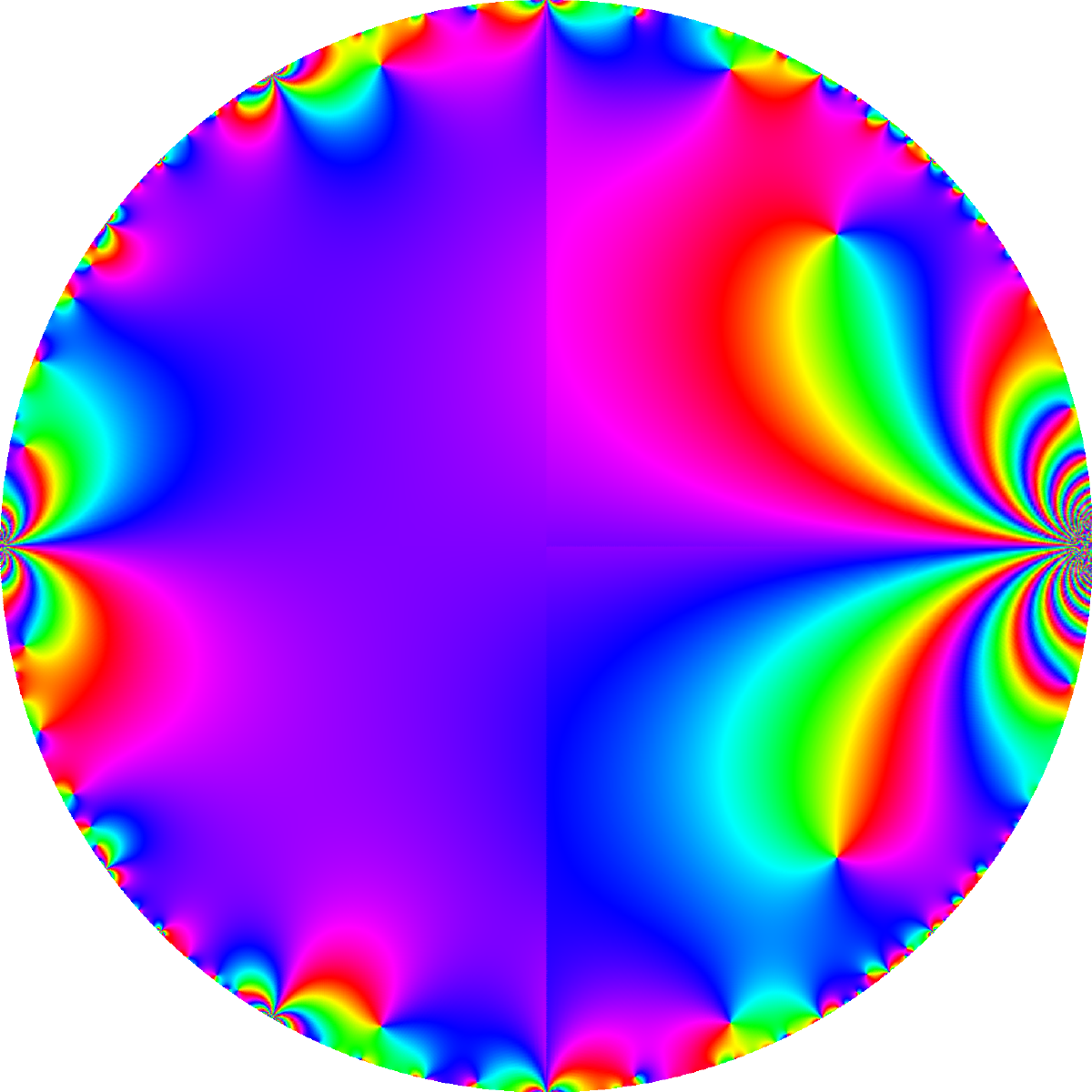}}
        \subfloat[$B^2(x,y)$]
        {\label{baker2}\includegraphics[width=0.35\linewidth]{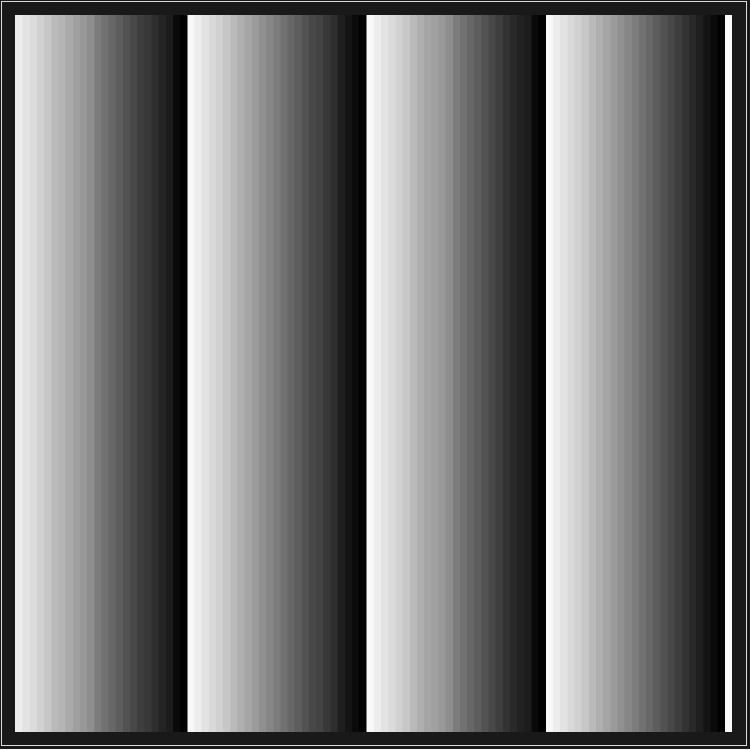}}
        \\
        \subfloat[$T_8Z(\tau)$]{\label{hecke8}\includegraphics[width=0.35\linewidth]{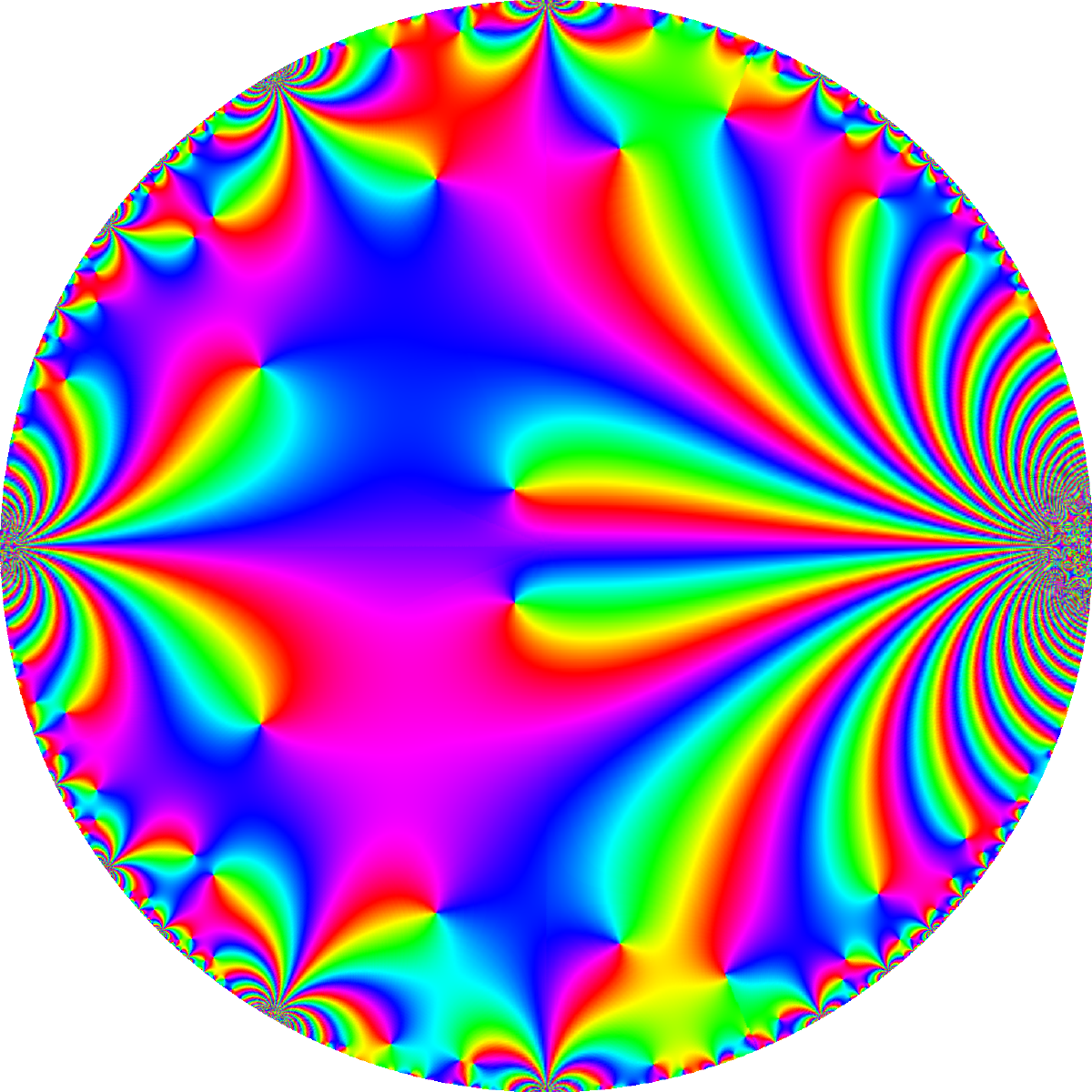}}
        \subfloat[$B^8(x,y)$]{\label{baker8}\includegraphics[width=0.35\linewidth]{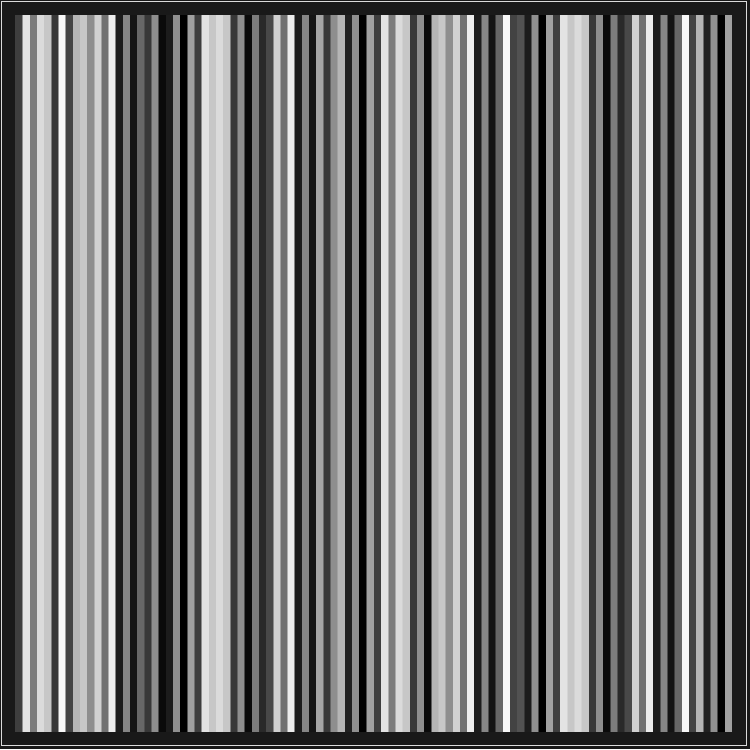}}
        \\
        \subfloat[$T_{16}Z(\tau)$]{\label{hecke16}\includegraphics[width=0.35\linewidth]{images/hecke16.pdf}}
        \subfloat[$B^{16}(x,y)$]{\label{baker16}\includegraphics[width=0.35\linewidth]{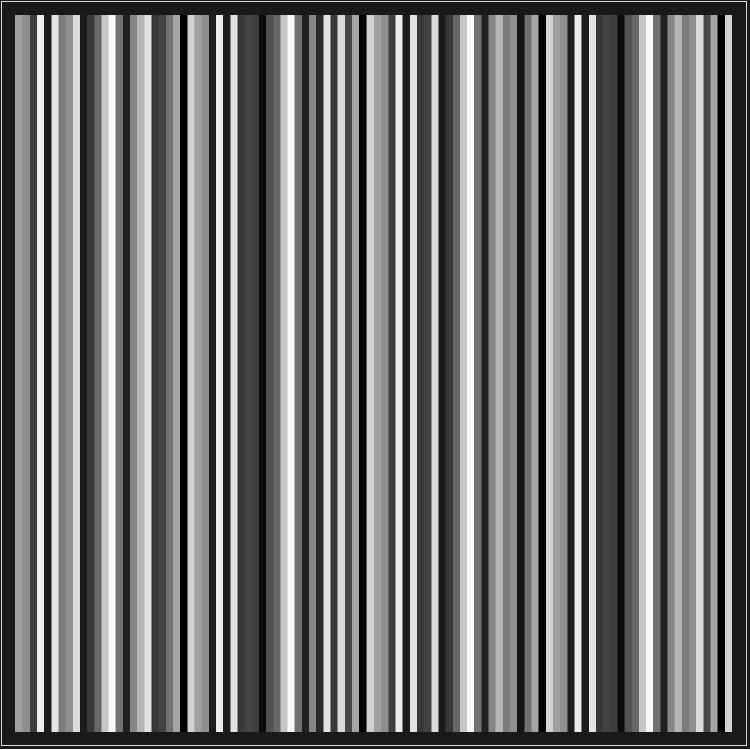}}
        \caption{An analogy between density plots of the action of Hecke operators $T_N$ on the torus partition function $Z(\tau)$ of the chiral boson, and the action of the Baker's map $B(x,y)$ (a classical chaotic map) on a gray square gradient.}
        \label{heckebaker}
    \end{figure}

\subsubsection*{An ergodic interpretation}
    Hecke equidistribution may have an ergodic interpretation, as it bears a resemblance to an ergodic theorem. For instance, the \textit{von Neumann ergodic theorem} states that any isometric operator $U$ acting on an element $h$ of a Hilbert space $\mathcal{H}$ admits the following limit
    \es{}{
    \lim_{N\to\infty}\frac{1}{N}\sum_{k=0}^{N-1}U^k h = \bar{h},
    }
    where $\bar{h}$ is the projection of $h$ onto the subspace of $U$-invariant states \cite{von_neumann_proof_1932}. This at least bears a bit of resemblance to \eqref{equidistribution}, in the sense that we are taking the limit of large $N$ for a sum like $\sum_{k=0}^{N}$ \footnote{We thank Nima Lashkari for pointing out this similarity.}. The only difference seems to be that instead of an isometric operator $U$ acting on a Hilbert space element $h\in\mathcal{H}$, we have fractional transformations $\tau\to\frac{N\tau+bd}{d^2}$ acting on the partition function $Z(\tau)$, where equidistribution occurs for the part $Z_{\text{spec}}\in L^2(\mathcal{F})$.
    \par
    Indeed, one may in fact \textit{prove} the $SL(2,\mathbb{Z})$ Hecke equidistribution theorem using ergodic theoretic techniques, as one of the authors of the original proof \cite{clozel_hecke_2001} went on to do in \cite{eskin_ergodic_2006}. This potential connection to ergodic theory is enticing, as there has been a great deal of work relating ergodic theory and von Neumann algebras to holography and bulk emergence \cite{gesteau_emergent_2023, furuya_information_2023, leutheusser_emergent_2023}. In particular, the emergence of type III$_1$ factors in the context of the Bost-Connes system \cite{bost_hecke_1995, neshveyev_von_2009},\footnote{Also known as a ``primon gas" in some physics literature \cite{hartnoll_conformal_2025, hartnoll_wheeler-dewitt_2023}} which forms a second quantized model with the Riemann zeta function as its partition function, may mean that similar phenomena may be found in holographic contexts, using recently developed arithmetic tools for CFT$_2$ \cite{perlmutter_l-function_2025}. \footnote{We thank Elliott Gesteau, Yasin Alam, and Daniel Murphy for pointing out the potential connection to ergodic theory and similarity to Bost-Connes systems.}

\acknowledgments

NC would like to thank Yasin Alam, Suzanne Bintanja, Hardik Bohra, Anatoly Dymarsky, Loki Lin, Daniel Murphy, Liza Rozenberg, and especially their advisor Alfred Shapere for extensive comments on the draft and helpful discussions. We thank the organizers of the 2025 workshop ``Quantum Information and Quantum Gravity" at the Perimeter Institute for Theoretical Physics during which many of these discussions took place. An early version of this work was presented at the workshop ``Quantum Gravity in Duluth" at the University of Minnesota, Duluth, and the 2025 Great Lakes Mathematical Physics meeting, at the University of Kentucky.

\appendix

\section{Orbifold phase transition}\label{phase}
As mentioned in section \ref{cyc}, we would like to see that the large $c$ vacuum phase transition that occurs for 2d CFTs is preserved under orbifolding \cite{keller_phase_2011}. In other words, we would like to see that taking the large $N$ limit for permutation orbifolds reproduces the vacuum phase transition structure of the BTZ black hole. To do this, we will summarize the argument in \cite{haehl_permutation_2015}, but particularly with an eye towards the action of Hecke operators. 
\par
To begin, we'll recall our orbifold partition functions, \eqref{snorb} and \eqref{sqfcyc}. Our strategy will be to show that in the large $N$ limit these reproduce the phase structure of the BTZ black hole, i.e. the phase structure of a large $c$ CFT at large and small temperature \cite{keller_phase_2011}:
\es{BTZ}{
Z(\beta) = \begin{cases}
    e^{\frac{c\beta}{12}} & \beta\gg1
    \\
    e^{\frac{c}{12}\frac{4\pi^2}{\beta^2}} & \beta\ll1
\end{cases}
}
This is indeed the Hawking-Page transition \cite{hawking_thermodynamics_1983} of a BTZ black hole, which tells us that the large $c$ limit invoked above is the semiclassical limit to be taken in gravity. This feature should be universal for any unitary 2d CFT in the large $c$ limit, so it should arise naturally for orbifold theories when we take the large $c=Nc_0$ limit, where $c_0$ is the central charge of the seed theory.
\par 
\subsubsection*{Cyclic product orbifold}

Recall the cyclic orbifold partition function \eqref{sqfcyc}. We will try to understand the large $N$ behavior of this partition function\footnote{In \cite{haehl_permutation_2015}, they only considered the prime $N$ case, but for us, the square-free Hecke operators will allow us to see the phase transition for general $N$.} by inspecting the large $N$ and small and large $\beta$ behavior of terms in the Hecke sum. First, we will restrict to the case of the rectangular torus, with $\tau=\frac{i\beta}{2\pi}$, so that our partition function will only be in terms of $\beta$ to see the HP transition as above. We will start with the low temperature ($\beta\gg1$) limit, where the vacuum contribution dominates the (nonchiral $c=\bar{c}$) partition function:
\es{}{
\left.Z(\tau)\right|_{\beta\gg1} \sim q^{\frac{-c}{24}}\bar{q}^{\frac{-\bar{c}}{24}} = e^{\frac{\beta c}{12}}.
}
% From here, we will inspect the vacuum contribution to the low temperature cyclic product orbifold partition function. 
% \es{}{
% \left.Z_{N,\mathbb{Z}}(\tau)\right|_{\beta\gg1} = \sum_{d|N}\frac{\phi(N/d)}{N}\sum_{b}T_{N/d}(e^{cd\beta/12}) - \sum_{d|N}\frac{\phi(N/d)}{d}\sum_{\substack{a\in\mathbb{Z}-1 \\ a^2|\frac{N}{d}}} \frac{1}{a^2}T^{sf}_{\frac{N}{da^2}}(x^{cd\beta/12}),
% }
% where we have expanded out the square divisor subtraction in the definition of the square-free Hecke operator \eqref{heckesquare}.
% Note that Hecke operators technically are supposed to act on modular forms (or functions at the very least), and the vacuum term is not modular invariant at all. However, we may still \textit{treat} this as behaving like a modular invariant quantity, since it was before we took the low temperature limit in the first place. Regardless, we can still take the Hecke sum anyway, since Hecke operators are just a way of acting on $\tau$, and do not necessarily \textit{need} to act on modular invariant functions (this assumption comes from assuming that the image, however, is modular invariant). Evaluating the Hecke sums explicitly on the vacuum contribution before substituting in $\tau=i\beta/2\pi$ gives
% \es{}{
% T_N(e^{cd\beta/12}) = \frac{1}{N}\sum_{s|N}s e^{\beta cdN/12s^2}.
% }
Similarly, taking the low temperature limit of \eqref{sqfcyc} after writing out the Hecke sums yields:
\es{}{
\left.Z_{N,\mathbb{Z}}(\tau)\right|_{\beta\gg1} = \sum_{d|N}\frac{\phi(N/d)}{N}\left[ \sum_{s|\frac{N}{d}}\left( se^{\beta N s^2 \frac{c}{12}} - \sum_{a^2|\frac{N}{d}}se^{\beta N s^2\frac{c}{12a^2}} \right) \right].
}
To handle the square divisor sum, we will assume that there is a single square divisor of $N$, and see what conclusions we can make in the case that there are more. For both terms, this will be the first place that we may make the large $N$ approximation: in the divisor sum over $s|\frac{N}{d}$, the $s=1$ term dominates, since it appears in the denominator in the exponential. So we can immediately rewrite for both terms for this reason:
\es{}{
\left.Z_{N,\mathbb{Z}}(\tau)\right|_{\beta\gg1} \sim \sum_{d|N}\frac{\phi(N/d)}{N}\left(  e^{\beta N \frac{c}{12}} - \sum_{a^2|\frac{N}{d}}e^{\beta N \frac{c}{12a^2}}  \right).
}
At this point, we know what we might expect: since the low temperature behavior is just the first exponential term. First of all, for the same reason as the previous step, the first term will exponentially dominate the second, since the second term, for any square divisor, will have an extra $a^2$ in the denominator of the exponential, and the square divisor term will not dominate for any square divisors (so this applies for general $N$). We're then left with 
\es{}{
\left.Z_{N,\mathbb{Z}}(\tau)\right|_{\beta\gg1} \sim \sum_{d|N}\frac{\phi(N/d)}{N} e^{\beta N \frac{c}{12}}.
}
Why should the prefactor be 1? We have the fact due to Gauss that $\sum_{d|N}\phi(d) = N$. This looks \textit{almost} like what we want. The last step is to notice that a sum over divisors $\sum_{d|N}f(d)$ is the same as the sum over ``divided out divisors": $\sum_{d|N}f(N/d)$, since both sums end up summing over the same divisors, just in the opposite order. So in the low temperature limit, we get the behavior we expect:
\es{lowt}{
\left.Z_{N,\mathbb{Z}}(\tau)\right|_{\beta\gg1} \underset{N\to\infty}{\sim} e^{\beta N\frac{c}{12}}.
}
What about\textit{ high temperature}? Modular invariance would imply that we can simply take the S-transformation of \eqref{lowt}, and we would indeed be done, but we may also proceed with an explicit calculation. Like in \cite{haehl_permutation_2015}, we can start with inspecting the terms in the Hecke sum $Z(\frac{N\tau+bd}{d^2})$. For $b=0$, we can get there by just taking an $S$-transformation and taking the low temperature limit again to just use the vacuum contribution:
\es{}{
Z\left(\frac{N\tau}{d^2}\right) \overset{S\text{-transform}}{=} Z\left(\frac{-d^2}{N\tau}\right) = e^{2\pi i(-d^2/N\tau)}e^{-2\pi i(-d^2/N\bar{\tau})} = e^{\frac{4\pi^2d^2}{N\beta}\frac{c}{12}}.
}
Next, for $b=1$, we may take similar steps, but with a little more manipulation:
\es{}{
Z\left( \frac{N\tau+d}{d^2} \right)  = Z\left( \frac{-d^2}{N\tau+d} \right)  = Z\left( \frac{Nd^2\tau}{N^2|\tau|^2+d^2} - \frac{d^3}{N^2|\tau|^2+d^2} \right).
}
Next, we'll do some rewriting to be able to use the large $N$ limit, and note that the second term is purely real, so it will not contribute for a nonchiral theory on a rectangular torus. Then,
\es{}{
Z\left( \frac{N\tau+d}{d^2} \right) = Z\left( \frac{\tau d^2/N}{|\tau|^2+d^2/N^2} \right) \sim Z(N\tau) \overset{S\text{-transform}}{=} Z\left(\frac{-1}{N\tau}\right)=e^{\frac{4\pi^2}{N\beta}\frac{c}{12}}
}
where we have used that $|\tau|\ll d/N$, since we are in the high temperature limit. Finally, the last nail in the coffin on the $b>0$ terms will be the fact that this is the same as the expression we had for the $b=0$ term, but without the $d^2$ in the exponent. Thus, the $b=0$ terms exponentially dominate the $b>0$ terms. So the large $N$ Hecke operator acts in the high temperature limit as:
\es{}{
\left.T_NZ(\tau)\right|_{\beta\ll1} \sim \frac{1}{N}\sum_{d|N}e^{\frac{4\pi^2d^2}{N\beta}\frac{c}{12}},
}
and this sum is now dominated by $d=N$. Finally we may obtain the high temperature vacuum contribution to the cyclic orbifold partition function:
\es{}{
\left.Z_{N,\mathbb{Z}}(\tau)\right|_{\beta\ll1}=\sum_{d|N}\frac{\phi(N/d)}{N}\left( e^{\frac{4\pi^2}{\beta}\frac{cN}{12}} \right), 
}
where as the square divisor term gets dominated for the same reason as before, and we have the Hawking-Page transition for cyclic product orbifold theories for any $N$:
\es{}{
\lim_{N\to\infty}Z_{N,\mathbb{Z}}(\tau) = 
\begin{cases}
    e^{\frac{cN\beta}{12}} & \beta\gg1
    \\
    e^{\frac{cN}{12}\frac{4\pi^2}{\beta}} & \beta\ll1
\end{cases}.
}
\subsubsection*{Symmetric product orbifold}
Finally, we may briefly use similar arguments for the symmetric product orbifold partition function. We may first use the low temperature action of the Hecke operator in \eqref{snorb}:
\es{}{
\left.Z_{N,S}(\tau)\right|_{\beta\gg1} = \sum_{\{ m_k \}}\prod_{k=1}^{N}\frac{1}{m_k!}\frac{1}{k^{m_k}}e^{\frac{\beta cm_k}{12}}.
}
With a little rewriting, we can use the number theoretic facts that $\sum_{k=1}^{N}m_k=N$ and $\sum_{\{m_k\}} \prod_{k=1}^{N}\frac{1}{k^{m_k}m_k!}=1$ to obtain:
\es{}{
\left.Z_{N,S}(\tau)\right|_{\beta\gg1} = \sum_{\{ m_k \}}\frac{\prod_{k=1}^{N}e^{\frac{\beta cm_k}{12}}}{\prod_{k=1}^{N}m_k!k^{m_k}} = e^{\frac{cN\beta}{12}}.
}
For high temperature, the argument goes through almost identically to obtain the final phase structure:
\es{}{
\left.Z_{N,S}(\tau)\right|_{\beta\gg1} = 
\begin{cases}
    e^{\frac{cN\beta}{12}} & \beta\gg1
    \\
    e^{\frac{cN}{12}\frac{4\pi^2}{\beta^2}} & \beta\ll1
\end{cases}.
}

% \section{A CFT trace formula}\label{trace}
% \import{Sections/}{trace}

\section{Relationship between cyclic product orbifolds and R\'enyi entropy}\label{renyi}
In section \ref{cyc}, we briefly mentioned that computing a cyclic product orbifold torus partition function is related to computing R\'enyi entropy. Here we review why this is, from the perspective of twist operators. 
\par
When we tensor together $n$ theories and gauge a discrete global symmetry that arises, we introduce operators that break the discrete symmetry. These operators are known as \textit{twist operators}; they account for the fact that, before gauging, we were free to shift around any operator $\mathcal{O}_n(z,\bar{z})$ in an $n^{\text{th}}$ copy of the seed theory to any of the other copies by moving the operator through the branch points that connect the replica sheets. These twist operators in some cases may be calculated explicitly in terms of ordinary states in the seed theory, e.g. \cite{takayanagi_free_2022}. When we gauge the symmetry the operator $\mathcal{O}_n(z,\bar{z})$ becomes 
\es{}{
\mathcal{O}_n(z,\bar{z}) = \tilde{\sigma}_n(z,\bar{z})\mathcal{O}_1(z,\bar{z})\sigma_n(z,\bar{z}), 
}
where $\sigma_n$ is the $n^{\text{th}}$ twist operator that transports the operator $\mathcal{O}_1$ from the original theory to the $n^{\text{th}}$ replica sheet, and $\tilde{\sigma}$ does the same in the opposite direction. These operators $\mathcal{O}_n$ are of conformal dimension $H_n=\overline{H}_n=\frac{c}{24}\left(n-\frac{1}{n}\right)$, where $c$ is the central charge of the seed theory. Before gauging, $\mathcal{O}_n(z,\bar{z}) = \mathcal{O}_1(z,\bar{z})$, whereas after, the new operator picks up the twist operators. 
\par 
One alternative way of computing the torus partition function of a cyclic product orbifold theory is by computing a two-point correlation function of twist operators on the sphere, where inserting cuts changes the topology to a torus \cite{takayanagi_free_2022,asplund_holographic_2015}:
\es{znrenyi}{
Z_{2,\mathbb{Z}}(\tau) \sim \langle \sigma_2(z_1)\tilde{\sigma}_2(w_1) \rangle,
}
where the twist operators are inserted at the edges of the interval $A=[z_1,w_1]$, and they create the cuts such that a $2\pi$ rotation about a cut brings an operator to the next sheet in the surface. This insertion of twist operators is analogous to the insertion of operators that twist the torus boundary conditions as in \eqref{zg}.
\par
Now, some comments are in order as to how this all relates to R\'enyi entropy. Among the most famous quantities to compute in quantum field theory is the R\'enyi entanglement entropy:
\es{}{
S^{(n)}_A = \Tr \rho_A^n,
}
where $\rho_A=\Tr_{A^c}\rho$ is the reduced density matrix associated to the geometric subregion $A$, where the complement is traced over to form the reduced density matrix. R\'enyi entropy has enjoyed the limelight in QFT in many ways, recently playing a role in the information paradox \cite{faulkner_quantum_2013, lewkowycz_generalized_2013, calabrese_entanglement_2004, renner_black_2021}. The $n^{\text{th}}$ R\'enyi entropy may be calculated as a stepping stone to calculating the \textit{von Neumann entropy}:
\es{}{
S_A = -\Tr \rho_A \log \rho_A = \lim_{n\to1}\frac{1}{n-1}\log S^{(n)}_A.
}
The von Neumann entropy is of interest for many reasons, but perhaps the most relevant for holography is the application in the \textit{Ryu-Takayanagi} proposal \cite{ryu_holographic_2006}, where boundary von Neumann entanglement entropy corresponds to the area of a bulk minimal surface. Since minimal surfaces are of pure geometric origin, their relationship to boundary entanglement entropy opens the door to probing bulk geodesics in the emergence of a bulk holographic spacetime. When we compute a \textit{replica} partition function in a 2d CFT seed theory in the vacuum state, we are computing
\es{}{
\Tr \rho^n_A = \langle \sigma_n(z_1)\tilde{\sigma}_n(w_1)\rangle,
}
and for $n=2$, this is of course \eqref{znrenyi}.

\bibliographystyle{JHEP}
\bibliography{references.bib}

\end{document}